\documentclass[%
reprint,
superscriptaddress,
amsmath,amssymb,
aps,
prb
]{revtex4-2}

\usepackage{amssymb,epsfig,amsmath,amsfonts}
\usepackage{graphicx}
\usepackage{dcolumn} %
\usepackage{bm,bbm}
\usepackage[utf8]{inputenc}     %
\usepackage[T1]{fontenc}        %
\usepackage{framed}
\usepackage{bold-extra}
\usepackage{tikz}
\definecolor{darkcopper}{RGB}{112,62,40}

\newcommand {\Evec}{{\mathbf{E}}}

\newcommand {\Bvec}{{\mathbf{B}}}
\newcommand {\Hvec}{{\mathbf{H}}}
\newcommand{\Heff}{{\mathbf{H}}_\text{eff}}

\newcommand {\Mvec}{{\mathbf{M}}}

\newcommand {\varphivec} {\boldsymbol{\varphi}}

\newcommand {\Jvec}{{\mathbf{J}}}

\newcommand {\Msat}{\text{M}_\text{s}}
\newcommand {\mvec}{{\mathbf{m}}}

\newcommand {\lex}{l_\text{ex}}

\newcommand {\xvec}{{\mathbf{x}}}

\newcommand {\vvec}{{\mathbf{v}}}

\newcommand {\wvec}{{\mathbf{w}}}
\newcommand {\evec}{{\mathbf{e}}}
\newcommand {\hvec}{{\mathbf{h}}}

\newcommand {\nvec}{{\mathbf{n}}}

\newcommand {\uvec}{{\mathbf{u}}}

\newcommand {\Avec}{{\mathbf{A}}}

\newcommand {\Cvec}{{\mathbf{C}}}

\newenvironment{frcseries}{\fontfamily{frc}\selectfont}{}

\begin{document}

\title{Micromagnetic Lagrangian and Hamiltonian Normal-Mode Formulation of Cavity Magnon--Photon Interactions}
	
\author{Claudio Serpico}
\affiliation{Department of Electrical Engineering and Information Technologies,
	University of Naples Federico II, 80125 Naples, Italy}

\author{Marco Lembo}
\affiliation{Department of Electrical Engineering and Information Technologies,
	University of Naples Federico II, 80125 Naples, Italy}
\affiliation{KU Leuven, Faculty of Engineering Science, B-3000 Leuven, Belgium}
\affiliation{imec, Kapeldreef 75, B-3001 Leuven, Belgium}

\author{Giovanni Miano}
\affiliation{Department of Electrical Engineering and Information Technologies,
	University of Naples Federico II, 80125 Naples, Italy}

\begin{abstract}
We investigate magnon--photon interactions in a resonant cavity starting from
the coupled Maxwell and Landau--Lifshitz equations linearized about an
arbitrary micromagnetic ground state. By introducing an electric vector
potential and a suitable interaction Lagrangian, the formulation avoids the
cancellation of large magnetostatic contributions that arises in the
conventional magnetic-vector-potential description and becomes inaccurate
under modal truncation. The magnetization and cavity fields are expanded in
the normal modes of the uncoupled magnetic and electromagnetic subsystems,
with coupling coefficients given by overlap integrals of the corresponding
eigenfunctions. The resulting multimode Lagrangian has a noncanonical
structure due to the gyrotropic terms, and the corresponding Hamiltonian
formulation is obtained using the Faddeev--Jackiw approach. The radiative
magnon--photon interaction is encoded in the resulting Poisson-bracket
structure, while the cavity-induced modification of the magnetostatic field
enters explicitly in the Hamiltonian. The coupled system is diagonalized
through a generalized Bogoliubov transformation, yielding canonical polariton
variables that provide a natural basis for quantization. Application to a
perpendicularly magnetized YIG disk in a rectangular microwave cavity shows
hybridization between the cavity mode and several magnetostatic modes,
together with appreciable cavity-wall-induced frequency shifts when the
sample is placed close to the conducting walls.
\end{abstract}

\maketitle

\section{Introduction}
\label{sec:intro}

In magnetically ordered media, magnon--photon interactions give rise to
hybrid collective excitations known as magnetic polaritons
\cite{Mills1974}. These interactions originate from the coupling between
the electromagnetic field and the normal modes of magnetization dynamics,
including spin waves and magnetostatic modes
\cite{Akhiezer1968,Walker1957,StancilPrabhakar2009,Rezende2020}.
The classical theory of magnetic polaritons emerged during the 1940s
and 1950s from the study of microwave electrodynamics in ferrite materials
\cite{Hogan1952,Polder1949,SuhlWalker1954a,SuhlWalker1954b,SuhlWalker1954c},
within the broader development of microwave and guided-wave electrodynamics
\cite{Marcuvitz1951,SchwingerCED1}. These studies established a
self-consistent theoretical framework in which Maxwell's equations are coupled
to a magnetic constitutive relation described by the Polder tensor
\cite{Polder1949}. This approach provided the basis for the development of
numerous microwave devices, including isolators, circulators, gyrators,
phase shifters, and tunable filters
\cite{GurevichMelkov1996,Pozar2012}.

In recent decades, the emergence of cavity magnonics
\cite{BhoiKim2019,HarderJAP2021,Rameshti2022} has renewed interest in the study of 
magnon--photon interactions, motivated both by prospective applications in
hybrid quantum technologies and quantum information processing
\cite{ClerkNatPhys2020,LachanceQuirion2019,TabuchiScience2015,
	LachanceQuirionScience2020}
and by the study of non-Hermitian physics
in hybrid magnon--photon systems \cite{AshidaAdvPhys2020,HurstFlebus2022}.
A key development behind this revival was the realization of the
strong-coupling regime between collective spin excitations in ferromagnets
and microwave cavity photons. This regime was first predicted theoretically
\cite{SoykalFlatte2010,SoykalFlattePRB2010} and subsequently demonstrated
experimentally with yttrium iron garnet (YIG) in three-dimensional microwave
cavities \cite{Huebl2013,Tabuchi2014,Zhang2014,Goryachev2014,Flower2019} and superconducting
coplanar microwave resonators \cite{TabuchiScience2015, LachanceQuirionScience2020}.

A distinctive feature of much of modern cavity magnonics, in both the
classical and quantum regimes
\cite{SoykalFlatte2010,SoykalFlattePRB2010,BhoiKim2019,Kusminskiy2019,
	BoventerPRR2020,HarderJAP2021,Rameshti2022},
is the adoption of Hamiltonian descriptions of coupled magnon--photon
dynamics, rather than formulations based  on Maxwell's equations
supplemented by Polder-type constitutive relations.
One important reason is that a Hamiltonian formulation provides a natural
route toward a quantum description through canonical quantization \cite{CohenTannoudji1989}.
Within this framework, the cavity-photon Hamiltonian is well established
and is naturally expressed in terms of the electromagnetic normal modes of
the resonator \cite{Dutra2005}, whereas the magnetic and interaction
Hamiltonians are often introduced at the level of effective models
\cite{BhoiKim2019,HarderJAP2021,Rameshti2022}. Expressions of the
magnon--photon coupling as overlap integrals between cavity and
micromagnetic mode profiles are available
\cite{LachanceQuirion2019,Flower2019}, but they are typically obtained by
inserting the cavity field into a Zeeman-type interaction, without a
systematic treatment of the magnetostatic field, which enters both
Maxwell's equations and the micromagnetic effective field \cite{Yu2020,MartinezLosaDelRincon2026}.

The main result of this work is a self-consistent Lagrangian and Hamiltonian
formulation of cavity magnonics, derived from the coupled Maxwell and
Landau--Lifshitz equations linearized about a possibly nonuniform micromagnetic
ground state, together with the corresponding normal-mode description. The cavity
electromagnetic field is expanded in the eigenfunctions of the vector Laplacian
\cite{Marcuvitz1951,Pozar2012}, while the magnetic excitations are represented
by the eigenmodes of the linearized Landau--Lifshitz operator
\cite{dAquino_2009,SerpicoPRB2025,Valet2026}, whose Hamiltonian retains the
usual micromagnetic form underlying spin-wave and magnetostatic-mode theory
\cite{Akhiezer1968,Walker1957,StancilPrabhakar2009,Rezende2020}.

The main difficulty arises when the two subsystems are coupled. The
micromagnetic effective field contains a magnetostatic contribution
\cite{Brown1963} that is also included in the magnetic field governed by
Maxwell's equations and must therefore be removed to avoid double counting.
In a formulation based on the conventional magnetic vector potential, this
requires the cancellation of large magnetostatic contributions. Although
exact at the field level, this cancellation is poorly reproduced after modal
truncation because the magnetostatic field is localized around the magnetic
sample and generally requires a very large number of cavity modes for its
representation. This severely limits the construction of reduced-order
Hamiltonian models.

To overcome this difficulty, we formulate the electromagnetic dynamics in
terms of the vector potential associated with the cavity electric field and
choose the interaction Lagrangian so that the magnetostatic contribution is
eliminated before the modal expansion is performed. 
The resulting multimode
Lagrangian has a noncanonical structure due to the gyrotropic terms, and the corresponding Hamiltonian
formulation is obtained using the Faddeev--Jackiw approach
\cite{Faddeev1988}. The radiative magnon--photon interaction is encoded in
the resulting Poisson-bracket structure, whereas the cavity-induced
modification of the magnetostatic field enters explicitly in the Hamiltonian
and accounts for the shielding produced by the cavity walls
\cite{NashStandley1960,Srivastava1966}. A generalized Bogoliubov
transformation \cite{Colpa1978,vanHemmen1980} then yields the hybridized
magnon--photon normal modes.

The formulation naturally combines the normal modes of the two uncoupled
subsystems. The magnetic modes can be obtained using standard micromagnetic
methods and computational tools for ferromagnets of arbitrary geometry and nonuniform ground state
\cite{DonahuePorter1999,Grimsditch_2004b,Rivkin_2004,McMicheal_2005, Schrefl2007, dAquino_2009,Fischbacher2007Nmag, 
	Abert2013MagnumFE, Bruckner2019, dAquinoHertel2023}, while the electromagnetic modes of the empty
cavity can be determined numerically or analytically
\cite{Marcuvitz1951,Pozar2012}. The corresponding magnon--photon coupling
coefficients are given by overlap integrals of the magnetic and
electromagnetic eigenfunctions, providing a reduced multimode description of
cavity magnetic polaritons suitable for subsequent quantization.

The paper is organized as follows. In Sec.~\ref{sec:matter+field_lagrangian},
we introduce the fundamental field equations, discuss the micromagnetic ground 
state and its stability, and derive the coupled evolution equations linearized 
about this ground state. In Sec.~\ref{sec:L_and_H_of_the_coupled_problem}, we introduce 
the electric vector potential and present the Lagrangian and Hamiltonian field formulation
of the coupled problem. In Sec.~\ref{sec:normal_modes_non_int}, we introduce the normal modes of the
uncoupled electromagnetic and magnetic subsystems. Then, in what constitutes
the central section of the paper, Sec.~\ref{sec:finite_dim_NM_L_and_H}, we derive
the normal-mode Lagrangian and Hamiltonian of the coupled system and outline the
canonical quantization procedure. The formalism is finally applied in
Sec.~\ref{sec:application} to a representative cavity-magnonics configuration
involving a YIG disk in a rectangular microwave cavity. The conclusions are
presented in Sec.~\ref{sec:conclusions}. Technical details and complementary
derivations, including the alternative formulation based on the magnetic vector
potential and the cancellation issues it entails, are provided in the Appendices
and in the Supplemental Material (referred to as SM in the following).

\section{Fundamental Equations}
\label{sec:matter+field_lagrangian}

\subsection{Field equations}

\begin{figure}[tb]
	\centering
	\begin{tikzpicture}[
		x=0.82cm,y=0.82cm,
		line cap=round,line join=round,
		>=stealth,
		every node/.style={font=\small}
		]
		\path[draw=black,fill=gray!14,line width=0.8pt]
		(-0.10,0) -- (1.60,0) -- (1.60,0.72) -- (0.48,0.72)
		-- (0.48,3.28) -- (1.60,3.28) -- (1.60,4) -- (-0.10,4) -- cycle;
		\node at (1.00,3.62) {$V_\mathrm{dc}$};
		
		\draw[rounded corners=12pt,draw=darkcopper,fill=white,line width=1.5pt]
		(1.77,0.25) rectangle (7.39,3.75);
		\draw[rounded corners=8pt,draw=black,fill=gray!38,line width=0.8pt]
		(3.02,1.35) rectangle (4.62,2.65);
		\node at (3.82,2.00) {$V_\mathrm{M}$};
		\node at (6.95,0.72) {$V$};
		
		\draw[->,line width=0.8pt] (4.62,2.08) -- (5.30,2.08);
		\node at (5.02,2.34) {$\mathbf{n}_\mathrm{M}$};
		\draw[line width=0.6pt] (4.48,1.40) -- (5.06,1.02);
		\node[anchor=west] at (5.10,1.00) {$\partial V_\mathrm{M}$};
		
		\draw[->,line width=0.8pt] (7.39,2.72) -- (6.69,2.72);
		\node at (6.99,2.98) {$\mathbf{n}$};
		\draw[line width=0.6pt] (7.39,1.14) -- (7.99,0.82);
		\node[anchor=west] at (8.03,0.80) {$\partial V$};
	\end{tikzpicture}
\caption{Schematic geometry (not to scale). The cavity occupies the region
	$V$, while the magnetic body occupies the subregion
	$V_\mathrm{M}\subset V$. The region $V_\mathrm{dc}$ contains the stationary
	sources of the applied dc field. The vector $\mathbf{n}$ is the inward unit
	normal on $\partial V$, while $\mathbf{n}_\mathrm{M}$ is the outward unit
	normal on $\partial V_\mathrm{M}$.}
	\label{fig:regions_and_boundaries}
\end{figure}
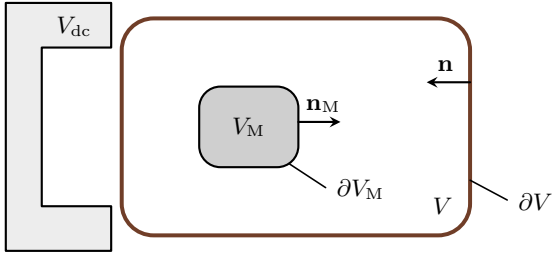

We consider a nonconducting ferromagnetic body placed inside a cavity bounded by
highly conducting walls. The dielectric polarization of the magnetic body is
assumed to be negligible, so that its electromagnetic response is described
solely through the magnetization field $\Mvec$. The cavity occupies a simply
connected region $V$, while the ferromagnetic body occupies a subregion
$V_\mathrm{M}\subset V$ (see Fig.~\ref{fig:regions_and_boundaries}).

We study the excitations arising from the coupling, within the region $V$,
between the electromagnetic fields $(\Evec,\Hvec)$ and the magnetization
$\Mvec$, for small deviations of $\Mvec$ from a stable micromagnetic
equilibrium configuration $\Mvec_0$. The equilibrium is established under the
action of the static magnetic field generated by a prescribed current density
$\Jvec_\text{dc}$ confined to a region $V_\text{dc}$. Since cavity walls do not shield static magnetic fields, the fields
generated by $\Jvec_\text{dc}$ and $\Mvec_0$ extend throughout space and decay
at infinity. The region $V_\text{dc}$ is assumed to lie outside the cavity
region $V$ (see Fig.~\ref{fig:regions_and_boundaries}).

The time evolution of the electromagnetic fields in $V$ is governed by Maxwell's
equations:
\begin{align}
	\label{eq:MaxwellrotE_f}
	& \nabla \times \Evec = - \mu_0\partial_t \Hvec
	-\mu_0\partial_t \Mvec \, , \\[4pt]
	\label{eq:MaxwellrotB_f}
	& \nabla \times \Hvec = \varepsilon_0 \, \partial_t \Evec \, , \\[4pt]
	\label{eq:MaxwelldivE_f}
	& \nabla \cdot \Evec = 0 \, , \\[4pt]
	\label{eq:MaxwelldivB_f}
	& \nabla \cdot \Hvec = -\nabla \cdot \Mvec \, ,
\end{align}
where we have adopted the Coulomb description of magnetic 
dipoles \cite{Bobbio1999,FanoChuAdler1960}.

The prescribed current density
$\Jvec_\text{dc}=\Jvec_\text{dc}(\xvec)$ generates the static magnetic field
\begin{align}
	\label{eq:static_B_field_dc_f}
	\Hvec_\text{dc}(\xvec)
	=
	\frac{1}{4\pi}
	\int_{V_\text{dc}}
	\frac{
		\Jvec_\text{dc}(\xvec^\prime)
		\times
		(\xvec-\xvec^\prime)
	}{
		|\xvec-\xvec^\prime|^3
	}
	\, dV_{\xvec^\prime} \, ,
\end{align}
which extends throughout space.

On the boundary of the magnetic body, $\partial V_\text{M}$, the field $\Hvec$ satisfies the interface conditions
\begin{align}
	\label{eq:interface_cond_1}
	& \nvec_\text{M} \,\cdot\, \left( \Hvec^+ - \Hvec^- \right)
	= \nvec_\text{M} \cdot \Mvec \, , \\
	\label{eq:interface_cond_2}
	& \nvec_\text{M} \times \left( \Hvec^+ - \Hvec^- \right)
	= 0 \, ,
\end{align}
where $\nvec_\text{M}$ denotes the outward unit normal to $\partial V_\text{M}$, while $\Hvec^+$ and $\Hvec^-$ denote the limiting values of $\Hvec$ on the exterior and interior sides of $\partial V_\text{M}$, respectively.

Maxwell's equations~\eqref{eq:MaxwellrotE_f}--\eqref{eq:MaxwelldivB_f}, \eqref{eq:interface_cond_1}--\eqref{eq:interface_cond_2} are complemented by
the constitutive relation governing the magnetization dynamics,
\begin{align}
	\label{eq:general_VLL_f}
	\partial_t \Mvec
	= \gamma \, \Mvec \times \left( \delta_\Mvec F - \mu_0\Hvec \right) ,
\end{align}
where $\gamma/ (2\pi) \approx 28.03\, \text{GHz}/\text{T}$
is the gyromagnetic ratio, and $F[\Mvec]$ is the micromagnetic Helmholtz
free-energy functional defined as
\begin{align}
	\label{eq:Micromagnetic_Energy_f}
	F[\Mvec]
	= \int_{V_\text{M}} \!\left[ A (\nabla \mvec)^2 + f_\text{an}(\mvec) \right] dV \, ,
\end{align}
with $\mvec=\Mvec/\Msat$ the unit magnetization vector. The quantity $\delta_{\Mvec}F$ denotes the
variational derivative of $F$ with respect to $\Mvec$ and is given by
\begin{align}
	\delta_\Mvec F
	= - {2A}\Msat^{-1} \nabla^2 \mvec
	+ \Msat^{-1} \partial_{\mvec} f_\text{an} \, ,
\end{align}
where $\partial_\mvec$ denotes the gradient with respect to $\mvec$.

The first term in eq.~\eqref{eq:Micromagnetic_Energy_f} represents the
exchange contribution, with
$(\nabla \mvec)^2 = \sum_{h,k} (\partial_h m_k)^2$ and
$A=\mu_0\Msat^2 l_\text{ex}^2/2$, where $l_\text{ex}$ is the exchange
length, typically of the order of tens of nanometers. The function
$f_\text{an}(\mvec)$ denotes the anisotropy energy density. In the uniaxial
case commonly considered in micromagnetics, it takes the form
$f_\text{an}(\mvec)= -K_1 (\mvec \cdot \evec_\text{an})^2$, where $K_1$
is the anisotropy constant and $\evec_\text{an}$ is the unit vector along
the anisotropy axis.

Equation~\eqref{eq:general_VLL_f} is the Landau--Lifshitz (LL) equation \cite{LandauLifshitz1935,Brown1963}, 
which is nonlinear, anisotropic, and dispersive in both space and time. 
Owing to its torque structure, the magnitude of the magnetization is conserved throughout the evolution. Accordingly, the magnetization field satisfies the micromagnetic constraint
\begin{align}
	\label{eq:micro_constraint_f}
	|\Mvec(\xvec, t)| = \Msat \, ,
\end{align}
where $\Msat$ is the saturation magnetization of the ferromagnet.

The magnetization is also required to satisfy the boundary condition
\begin{align}
	\label{eq:exchange_CC_0_f}
	\nvec_\text{M} \cdot \nabla \Mvec = 0
	\qquad \text{on } \partial V_\text{M} \, ,
\end{align}
where $\nvec_\text{M}$ is the outward unit normal to $\partial V_\text{M}$.
This exchange boundary condition arises naturally from the variational treatment
of the exchange term in the micromagnetic energy functional~\eqref{eq:Micromagnetic_Energy_f}.

In the following, the highly conducting cavity walls are treated in the
ideal limit of perfect conductivity. Accordingly, the electric field satisfies
the boundary condition
\begin{align}
	\label{eq:E_boundary_cond_f}
	\nvec \times \Evec = 0
	\qquad \text{on } \partial V \, ,
\end{align}
where $\nvec$ denotes the inward unit normal to $\partial V$.

Writing the magnetic constitutive relation in the form of
eq.~\eqref{eq:general_VLL_f} is customary in formulations based on local
constitutive relations, such as those employing the Polder tensor
\cite{Polder1949,Pozar2012}. In micromagnetic treatments, however, the
LL equation for a free magnetic body is usually written in terms
of the effective field
\begin{align}
	\nonumber
	\mu_0\Heff[\Mvec] &= \\
	&-\delta_\Mvec G
	= -\delta_\Mvec F + \mu_0(\Hvec_\text{dc} + \Hvec_\Mvec[\Mvec]) \, .
	\label{eq:Heffqs0_f}
\end{align}
The functional $G$ in eq.~\eqref{eq:Heffqs0_f} is the standard
micromagnetic Gibbs free energy \cite{Brown1962,Brown1963},
\begin{align}
	\nonumber
	G[\Mvec] 	= & F[\Mvec] \\
	&- \int_{V_\text{M}}
	\!\left[
	\frac{\mu_0}{2} \Mvec \cdot \Hvec_\Mvec[\Mvec]
	+ \mu_0\Mvec \cdot \Hvec_\text{dc}
	\right] dV \, ,
	\label{eq:Gibbs_energy_qs_f}
\end{align}
where $F[\Mvec]$ is defined in eq.~\eqref{eq:Micromagnetic_Energy_f},
while $\Hvec_\Mvec[\Mvec]$ denotes the magnetostatic field generated by
$\Mvec$. Its explicit expression is
\begin{align}
	\nonumber
	\Hvec_\Mvec[\Mvec](\xvec)
	=& -\nabla \!\left[
	\frac{1}{4\pi}
	\int_{V_\text{M}}
	\frac{-\nabla'\!\cdot\!\Mvec(\xvec')}{|\xvec-\xvec'|} \, dV_{\xvec'}
	+ \right. \\[2pt]
	&\left. \frac{1}{4\pi}
	\oint_{\partial V_\text{M}}
	\frac{\nvec_\text{M}(\xvec')\!\cdot\!\Mvec(\xvec')}{|\xvec-\xvec'|} \, dS_{\xvec'}
	\right] \, ,
	\label{eq:Hm_operator_f}
\end{align}
which is the curl-free demagnetizing field generated in free space by
the volume and surface magnetic charges associated with $\Mvec$.

The LL equation~\eqref{eq:general_VLL_f} can then be rewritten in terms
of the effective field~\eqref{eq:Heffqs0_f} as
\begin{align}
	\nonumber
	\partial_t & \Mvec
	= \\
	&-\gamma\mu_0 \Mvec \times \left[
	\Heff[\Mvec]
	+ (\Hvec -\Hvec_\text{dc}
	- \Hvec_\Mvec[\Mvec])
	\right] .
	\label{eq:general_VLL_Gqs}
\end{align}
This is the form of the LL equation that will be used in what follows.
It is worth noting that, to avoid double counting of the static magnetic
contributions already included in $\Heff[\Mvec]$, only the difference
$\Hvec-(\Hvec_\text{dc}+\Hvec_\Mvec[\Mvec])$ enters as an additional field.

\subsection{Micromagnetic ground state}
\label{sec:linear_mag_dyn}

The magnetization dynamics in cavity magnonics and polariton physics
typically involves small oscillations of the magnetization around a
stationary \emph{micromagnetic ground state} $\Mvec_0$.
The stationary solution of Maxwell's equations,
eqs.~\eqref{eq:MaxwellrotE_f}--\eqref{eq:MaxwelldivB_f} and
\eqref{eq:interface_cond_1}--\eqref{eq:interface_cond_2},
is given by $\Evec_0=0$ and
\begin{align}
\Hvec_0=\Hvec_\text{dc}+\Hvec_\Mvec[\Mvec_0]
\label{eq:stationary_H0}\,.
\end{align}
Since $\Hvec_0$ is static, the cavity boundary does not affect
the determination of the stationary solution.

Substituting $\Mvec=\Mvec_0$ and  $\Hvec=\Hvec_0$ in 
eq.~\eqref{eq:general_VLL_Gqs}, we obtain \emph{Brown's equations}
\cite{Brown1962,Brown1963} for the micromagnetic equilibrium:
\begin{align}
	\label{eq:equilibium_cond_1_m_f}
	\Mvec_0 \times \Heff[\Mvec_0] = 0 \, ,
	\quad |\Mvec_0| = \Msat \, , \,\,\, \text{in } V_\text{M} \, ,
\end{align}
together with the boundary condition (see
eq.~\eqref{eq:exchange_CC_0_f})
\begin{align}
	\label{eq:equilibrium_cond_3_m_f}
	\nvec_\text{M} \cdot \nabla\, \Mvec_0 = 0
	\,\,\, \text{on } \partial V_\text{M} \, .
\end{align}

The first condition in Eq.~\eqref{eq:equilibium_cond_1_m_f} can be equivalently written as
\begin{align}
	\label{eq:equilibium_cond_1_m_b_f}
	\Heff[\Mvec_0] = h_i \Mvec_0 \, ,
\end{align}
where $h_i = h_i(\xvec)$ denotes the normalized equilibrium internal field.

The ground state is assumed to be a stable micromagnetic equilibrium, with
stability understood in the energetic sense: sufficiently small admissible
static perturbations must increase the micromagnetic Gibbs energy, so that
$\Mvec_0$ is a relative minimum of $G[\Mvec]$. To characterize this property,
we consider the second-order expansion of the Gibbs energy functional
$G[\Mvec]$ (see eq.~\eqref{eq:Gibbs_energy_qs_f}) about $\Mvec_0$, restricting
the variations of $\Mvec$ to the manifold $\mathcal{M}(V_\text{M})$
\cite{Brown1963}:
\begin{align}
	\nonumber
	\mathcal{M}(V_\text{M}) = & \left\{
	\Mvec(\xvec) :
	|\Mvec(\xvec)| = \Msat \,\, \text{in } V_\text{M} \,, \right.\\[-1pt]
	\label{eq:inf_dim_Manifold_M_bis}
	& \left.
	\nvec_\text{M} \cdot \nabla\, \Mvec = 0
	\,\, \text{on } \partial V_\text{M} \, , \;
	G[\Mvec]<\infty
	\right\} ,
\end{align}
that is, the set of magnetization fields with fixed magnitude $\Msat$,
finite Gibbs energy, and satisfying homogeneous Neumann boundary conditions
on $\partial V_\text{M}$.

A convenient local representation of fields in $\mathcal{M}(V_\text{M})$ is given by
\begin{equation}
	\label{eq:M_representatiom_app_0_f}
	\Mvec(\xvec,t)
	=
	\frac{\Mvec_0(\xvec)}{\Msat}
	\sqrt{\Msat^2-|\Mvec_1(\xvec,t)|^2}
	+\Mvec_1(\xvec,t) \, ,
\end{equation}
where $\Mvec_1(\xvec,t)$ belongs to the linear space
\begin{align}
	\nonumber
	& \mathbb{H}^\text{sw}(V_\text{M}) =
	\left\{ 
	\vvec(\xvec) \, : \, 
	\vvec(\xvec) \cdot \Mvec_0(\xvec) = 0 
	\,\, \text{in } V_\text{M} \,, \right.\\
	\label{eq:spin-wave_space_f}
	& \quad \left.  
	\nvec_\text{M} \cdot \nabla \vvec = 0 
	\,\, \text{on } \partial V_\text{M}, \;
	\{\vvec,\nabla \vvec\} \in \mathbb{L}^2(V_\text{M})
	\right\} ,
\end{align}
whose elements are vector fields that are pointwise orthogonal to $\Mvec_0(\xvec)$, possess square-integrable components and gradients, and satisfy homogeneous Neumann boundary conditions.
It is readily verified that the field $\Mvec(\xvec,t)$ defined by Eq.~\eqref{eq:M_representatiom_app_0_f}, with $\Mvec_1 \in \mathbb{H}^\text{sw}(V_\text{M})$, belongs to $\mathcal{M}(V_\text{M})$. 
The representation \eqref{eq:M_representatiom_app_0_f} is one-to-one only for magnetization fields satisfying
$\Mvec(\xvec) \cdot \Mvec_0(\xvec) > 0$ in $V_\text{M}$. This restriction is not limiting when considering small deviations from equilibrium.

The space $\mathbb{H}^\text{sw}(V_\text{M})$ can be interpreted as the linear space of \emph{spin-wave fields} (as indicated by the superscript), representing small-amplitude, low-energy excitations of the magnetization about the ground state.
In anticipation of the generalizations developed later in this manuscript, the vector fields in $\mathbb{H}^\text{sw}(V_\text{M})$ are taken to be complex-valued.

Expanding eq.~\eqref{eq:M_representatiom_app_0_f} in powers of $\Mvec_1 / \Msat$, we obtain
\begin{equation}
	\label{eq:M_representatiom_expansion_0_f}
	\Mvec = \Mvec_0 + \Mvec_1 
	- \frac{1}{2} \Mvec_0 \left| \frac{\Mvec_1}{\Msat} \right|^2
	+ \mathcal{O}(3) \, ,
\end{equation}
where $\mathcal{O}(n)$ denotes terms of order $|\Mvec_1/\Msat|^n$ or higher.
Substituting  expansion \eqref{eq:M_representatiom_expansion_0_f} into eq.~\eqref{eq:Gibbs_energy_qs_f} and using condition~\eqref{eq:equilibium_cond_1_m_b_f}, we obtain
\begin{align}
	\label{eq:expansion_G_M1_f}
	G[\Mvec] = G[\Mvec_0] + \frac{\mu_0}{2} 
	\int_{V_\text{M}} \!\!\!\!\!
	\Mvec_1 \cdot \mathbbm{D}_0[\Mvec_1] \, dV 
	+  \mathcal{O}\!\left(3\right) \, ,
\end{align}
where
\begin{align}
	\nonumber
	&	\mathbbm{D}_0[\Mvec_1] =  \\[2pt]
	\label{eq:operator_D1_f}
	&  -\lex^2 \nabla^2 \Mvec_1 + {\mathbbm k}_0\cdot \Mvec_1
	- \Hvec_\Mvec[\Mvec_1]
	+ h_i \Mvec_1 \, ,
\end{align} 
and ${\mathbbm k}_0$ is  the normalized  Hessian of $f_\text{an}(\mvec)$
\begin{align}
	\label{eq:Hessian_f_0_f}
	[{\mathbbm k}_0]_{ij} 
	= \frac{1}{\mu_0 \Msat^2}
	\left. \frac{\partial^2 f_\text{an}}{\partial m_i \, \partial m_j} \right|_{\mvec_0} \, ,
\end{align}
computed for $\mvec = \mvec_0= \Mvec_0/\Msat$. 
The operator $\mathbbm{D}_0[\,\cdot\,]$ is defined on the linear space $\mathbb{H}^\text{sw}(V_\text{M})$ of spin-wave fields.
To obtain an operator that maps $\mathbb{H}^\text{sw}(V_\text{M})$ into itself, we introduce
\begin{align}
	\mathbbm{D}_{0\bot}[\vvec] 
	= \mathbbm{P}_{0\bot}(\xvec) \cdot \mathbbm{D}_{0}[\vvec] \, ,
	\label{eq:def_of_D0bot_f}
\end{align}
where $\mathbbm{P}_{0\bot}(\xvec)$ is the projection operator onto the local tangent plane orthogonal to $\mvec_0(\xvec)$, defined by
\begin{align}
	\label{eq:def_of_D0bot_f_1}
	\mathbbm{P}_{0\bot}(\xvec) \cdot \wvec(\xvec) 
	= \mvec_0(\xvec) \times \!\left[\wvec(\xvec)\times \mvec_0(\xvec)\right] \, ,
\end{align}
for any vector field $\wvec(\xvec)$.

In conclusion, the stability of the equilibrium is governed by the quadratic
functional
\begin{align}
	\label{eq:Hamiltonian_spin-waves}
	\mathcal{H}_\text{M}[\Mvec_1]
	=
	\frac{\mu_0}{2}
	\int_{V_\text{M}}
	\Mvec_1 \cdot \mathbbm{D}_{0\bot}[\Mvec_1] \, dV \, ,
\end{align}
associated with the linear operator $\mathbbm{D}_{0\bot}[\,\cdot\,]$ defined
above. This operator is real and Hermitian with respect to the standard
$\mathbb{L}^2$ inner product, namely
\begin{align}
	\label{eq:Hermitian_D0bot}
	\int_{V_\text{M}}
	\vvec^* \cdot \mathbbm{D}_{0\bot}[\wvec] \, dV
	=
	\int_{V_\text{M}}
	\wvec \cdot
	\left(\mathbbm{D}_{0\bot}[\vvec]\right)^* \, dV \, ,
\end{align}
for any $\vvec,\wvec \in \mathbb{H}^\text{sw}(V_\text{M})$, where `$^*$'
denotes complex conjugation. The equilibrium $\Mvec_0$ is a relative minimum
of $G[\Mvec]$ if
$\mathcal{H}_\text{M}[\Mvec_1]>0$ for every nonzero real-valued
$\Mvec_1 \in \mathbb{H}^\text{sw}(V_\text{M})$, or, equivalently, if
$\mathbbm{D}_{0\bot}[\,\cdot\,]$ is positive definite. This stability
condition will be assumed throughout the following analysis.

The functional $\mathcal{H}_\text{M}$ defined in eq.~\eqref{eq:Hamiltonian_spin-waves} plays
the important role of the free-space spin-wave Hamiltonian. It is useful to write it
explicitly as
\begin{align}
	\label{eq:operator_D1_f_energy0}
	\mathcal{H}_\text{M} [\Mvec_1]
	= F_0[\Mvec_1] - \int_{V_\text{M}}
	\frac{\mu_0}{2} \Mvec_1 \cdot \Hvec_\Mvec[\Mvec_1] \, dV \, ,
\end{align}
where
\begin{widetext}
	\begin{align}
		\label{eq:operator_D1_f_energy}
		F_0[\Mvec_1]
		= \int_{V_\text{M}} \left[
		\frac{\mu_0}{2} \lex^2 (\nabla \Mvec_1)^2
		+ \frac{\mu_0}{2}  \Mvec_1 \cdot \mathbbm{k}_0 \cdot \Mvec_1
		+ \frac{\mu_0}{2} h_i \, \Mvec_1^2
		\right] dV \, ,
	\end{align}
\end{widetext}
with $h_i$ defined in eq.~\eqref{eq:equilibium_cond_1_m_b_f}.

Equation~\eqref{eq:operator_D1_f_energy0} splits the free-space spin-wave energy
into a local part, $F_0[\Mvec_1]$, which collects the exchange, anisotropy, and
equilibrium internal-field contributions, and the magnetostatic self-energy of
$\Mvec_1$, which is nonlocal.

\subsection{Coupled linearized evolution equations}
\label{sec:Decomposition_of_B1}

We now consider the evolution problem defined by
eqs.~\eqref{eq:MaxwellrotE_f}--\eqref{eq:E_boundary_cond_f}
for small deviations from the static solution
$\Mvec=\Mvec_0$, $\Hvec=\Hvec_0$ (see eq.~\eqref{eq:stationary_H0}),
and $\Evec=\Evec_0=0$. The deviation of the magnetization from
equilibrium is described by the field $\Mvec_1$ through the
expansion~\eqref{eq:M_representatiom_expansion_0_f}, while the
first-order electromagnetic fields $\Hvec_1$ and $\Evec_1$ are introduced
through
\begin{align}
	\label{eq:decomp_em_field-lin}
	\Hvec = \Hvec_0 + \Hvec_1 \, , \quad \Evec = \Evec_1 \, .
\end{align}
Substituting these expressions into
eqs.~\eqref{eq:MaxwellrotE_f}--\eqref{eq:MaxwelldivB_f} and
\eqref{eq:interface_cond_1}--\eqref{eq:interface_cond_2},
we obtain the Maxwell equations for first-order quantities:
\begin{align}
	\label{eq:MaxwellrotE1_EB_coup_b}
	& \nabla \times \Evec_1 = -\mu_0\partial_t \Hvec_1 -\mu_0\partial_t \Mvec_1\, , \\
	\label{eq:MaxwellrotB1_EB_coup_b}
	& \nabla \times \Hvec_1 = \varepsilon_0 \partial_t \Evec_1\, ,\\
	\label{eq:MaxwelldivE1_EB_coup_b}
	& \nabla \cdot \Evec_1 = 0 \, ,\\
	\label{eq:MaxwelldivH1_EB_coup_b}
	& \nabla \cdot \Hvec_1 = -\nabla \cdot \Mvec_1 \, ,
\end{align}
in the cavity volume $V$, together with the interface conditions
\begin{align}
	\label{eq:interface_cond_1_on_H1}
	& \nvec_\text{M} \cdot \left( \Hvec_1^+ - \Hvec_1^- \right)
	= \nvec_\text{M} \cdot \Mvec_1 \, , \\
	\label{eq:interface_cond_2_on_H1}
	& \nvec_\text{M} \times \left( \Hvec_1^+ - \Hvec_1^- \right)
	= 0 \, ,
\end{align}
on the boundary $\partial V_\text{M}$.

Moreover, from eq.~\eqref{eq:E_boundary_cond_f}, we obtain the boundary condition
\begin{align}
	\label{eq:E_boundary_cond_E1}
	\nvec \times \Evec_1 = 0  \, ,
\end{align}
on $\partial V$. Combining
eqs.~\eqref{eq:MaxwellrotE1_EB_coup_b} and \eqref{eq:E_boundary_cond_E1},
we further obtain
\begin{align}
	\label{eq:E_boundary_cond_B1}
	\nvec \cdot \left( \Hvec_1 + \Mvec_1 \right ) = \nvec \cdot \Hvec_1 = 0 \, ,
\end{align}
on $\partial V$, where we used the fact that $\Mvec_1=0$ on $\partial V$.

The equations \eqref{eq:MaxwellrotE1_EB_coup_b}--\eqref{eq:E_boundary_cond_B1} above must be supplemented by the linearized LL equation,
which is obtained by substituting the expansions~\eqref{eq:M_representatiom_expansion_0_f}
and~\eqref{eq:expansion_G_M1_f} into eq.~\eqref{eq:general_VLL_Gqs}
and retaining only the terms linear in $\Mvec_1$:
\begin{align}
	\label{eq:general_VLL_Gqs_lin}
	\partial_t \Mvec_1
	=  \gamma \Mvec_0 \times \left[
	\delta_{\Mvec_1} \mathcal{H}_\text{M}
	- \mu_0(\Hvec_1-\Hvec_\Mvec[\Mvec_1])
	\right] \, ,
\end{align}
where $\mathcal{H}_\text{M}$ is the quadratic functional defined in Eqs.~\eqref{eq:operator_D1_f_energy0}--\eqref{eq:operator_D1_f_energy},
$\delta_{\Mvec_1} \mathcal{H}_\text{M} = \mu_0\mathbbm{D}_{0\bot}[\Mvec_1]$
(see eqs.~\eqref{eq:operator_D1_f}, \eqref{eq:def_of_D0bot_f}, \eqref{eq:def_of_D0bot_f_1}).
Equation~\eqref{eq:general_VLL_Gqs_lin} governs the dynamics of 
$\Mvec_1$ within the spin-wave space defined in eq.~\eqref{eq:spin-wave_space_f}.

In the static approximation and in the absence of the cavity, $\Hvec_1 = \Hvec_\Mvec[\Mvec_1]$, and eq.~\eqref{eq:general_VLL_Gqs_lin}
assumes the usual  noncanonical Hamiltonian form, with $\mathcal{H}_\text{M}$ playing
the role of the Hamiltonian. This limiting case describes purely magnonic
dynamics \cite{Rezende2020,Walker1957,StancilPrabhakar2009}.

Notice that $\Hvec_1$ admits the Helmholtz decomposition
(see Appendix~\ref{sec:Helmholtz_Decomposition})
\begin{align}
	\label{eq:B1R_field}
	\Hvec_1 = \Hvec_\text{1R} + \Hvec_{\Mvec}^V[\Mvec_1] \, ,
\end{align}
where $\Hvec_{\Mvec}^V[\Mvec_1]$ is the curl-free component generated by the magnetization, whereas $\Hvec_\text{1R}$ is the solenoidal component of $\Hvec_1$, with ``R'' standing for radiative. This decomposition separates the magnetostatic field determined by the magnetization dynamics from the independent dynamical degrees of freedom of the electromagnetic field.

Decomposition~\eqref{eq:B1R_field} is the magnetic counterpart of the familiar decomposition of the electric field used in the theory of dielectric polaritons~\cite{Kittel1987}.

The field $\Hvec_{\Mvec}^V[\Mvec_1]$ is defined as the solution of
\begin{align}
	\label{eq:Maxwell_for_B1M}
	& \nabla \times \Hvec_{\Mvec}^V = 0 \, , \\
	& \nabla \cdot \Hvec_{\Mvec}^V = -\nabla \cdot \Mvec_1 \, ,
\end{align}
in $V$, subject to the interface conditions
\begin{align}
	\label{eq:interface_cond_1_on_H1M}
	& \nvec_\text{M} \,\cdot\, \left( \Hvec_\Mvec^{V+} - \Hvec_\Mvec^{V-} \right)
	= \nvec_\text{M} \cdot \Mvec_1 \, , \\
	\label{eq:interface_cond_2_on_H1M}
	& \nvec_\text{M} \times \left( \Hvec_\Mvec^{V+} - \Hvec_\Mvec^{V-} \right)
	= 0 \, ,
\end{align}
on $\partial V_\text{M}$, together with the boundary condition
\begin{align}
	\label{eq:BC_on_BM^V}
	\nvec \cdot \Hvec_\Mvec^V = 0 \, ,
	\qquad
	\nvec \cdot \Bvec_\Mvec^V = 0
	\qquad \text{on } \partial V \, ,
\end{align}
where
$\Bvec_\Mvec^V[\Mvec_1]
= \mu_0\left(\Hvec_\Mvec^V[\Mvec_1]+\Mvec_1\right)$
is a solenoidal field in $V$. Since the magnetic body is entirely contained
within the cavity, $\Mvec_1=0$ on $\partial V$, so that the two boundary
conditions in~\eqref{eq:BC_on_BM^V} are equivalent.

Notice that $\Hvec_{\Mvec}^V[\Mvec_1]$ satisfies the same equations as $\Hvec_\Mvec[\Mvec_1]$, defined in eq.~\eqref{eq:Hm_operator_f}, but with different boundary conditions. In this respect, the superscript ``V'' is used for quantities that satisfy the appropriate cavity
boundary conditions and to distinguish them from their free-space counterparts. The field $-\mu_0\Hvec_{\Mvec}^V[\Mvec_1]$ can also be interpreted as the curl-free component in the
following Helmholtz decomposition
\begin{align}
	\mu_0\Mvec_1 = \Bvec_\Mvec^V - \mu_0\Hvec_\Mvec^V \, ,
	\label{eq:Helmholtz_decomp_M1_0}
\end{align}
of $\mu_0 \Mvec_1$ over the cavity region $V$, with the additional condition that, according to eq.~\eqref{eq:BC_on_BM^V}, the normal component of $\Bvec_\Mvec^V$ vanishes on the boundary $\partial V$ (see Appendix~\ref{sec:Helmholtz_Decomposition}).

From the above considerations,
eqs.~\eqref{eq:MaxwellrotE1_EB_coup_b}--\eqref{eq:interface_cond_2_on_H1} yield
\begin{align}
	\label{eq:MaxwellrotE1_EB_coup_bR}
	& \nabla \times \Evec_1 = -\mu_0\partial_t \Hvec_\text{1R} -\partial_t \Bvec_{\Mvec}^V[\Mvec_1]\, , \\
	\label{eq:MaxwellrotB1_EB_coup_bR}
	& \nabla \times \Hvec_\text{1R} =   \varepsilon_0 \partial_t \Evec_1  \, ,\\
	\label{eq:MaxwelldivE1_EB_coup_bR}
	& \nabla \cdot \Evec_1 = 0 \, ,\\
	\label{eq:MaxwelldivH1_EB_coup_bR}
	& \nabla \cdot \Hvec_\text{1R} = 0 \, ,
\end{align}
and the last equation is obtained by substituting the decomposition \eqref{eq:B1R_field} 
into eq. \eqref{eq:MaxwelldivH1_EB_coup_b}.  
Equations~\eqref{eq:E_boundary_cond_B1} and \eqref{eq:BC_on_BM^V} lead to the
following boundary condition
\begin{align}
	\label{eq:E_boundary_cond_B1R}
	\nvec \cdot \Hvec_\text{1R} = 0\, , \qquad \text{on} \quad \partial V \, .
\end{align}
Notice that $\Hvec_\text{1R}$ remains continuous across $\partial V_\text{M}$, since the discontinuity of the magnetic field is  
accounted for by the field $\Hvec_\Mvec^V[\Mvec_1]$ through eqs.~\eqref{eq:interface_cond_1_on_H1M} and \eqref{eq:interface_cond_2_on_H1M}.

Using the decomposition \eqref{eq:B1R_field}, the linearized LL equation \eqref{eq:general_VLL_Gqs_lin} can be rewritten as
\begin{align}
	\nonumber
	\partial_t \Mvec_1
	&=  \\
&	\gamma \Mvec_0 \times
	\left[
	\delta_{\Mvec_1} \mathcal{H}_\text{M}
	- \mu_0 \Hvec_\text{1R}
	-\mu_0\Hvec_{1\Mvec}^\Delta[\Mvec_1]
	\right] ,
	\label{eq:general_VLL_Gqs_lin_bis}
\end{align}
where 
\begin{align}
	\label{eq:def_of_H1delta}
	\Hvec_{1\Mvec}^\Delta[\Mvec_1] = \Hvec_{\Mvec}^V[\Mvec_1]-\Hvec_\Mvec[\Mvec_1]
\end{align}
is such that
\begin{align}
	& \nabla \cdot \Hvec_{1\Mvec}^\Delta = 0  \, ,  \quad  \nabla \times \Hvec_{1\Mvec}^\Delta = 0 \quad \text{in} \quad V \, ,\\
	\label{eq:BC_for_psi}
	& \nvec \cdot \Hvec_{1\Mvec}^\Delta = 
	-\nvec \cdot \Hvec_\Mvec[\Mvec_1]
	\quad \text{on} \quad \partial V \, .
\end{align}
It follows that the harmonic field  $\Hvec_{1\Mvec}^\Delta $ depends on the vector field  $\Mvec_1$ only through the boundary condition \eqref{eq:BC_for_psi}. The field $\Hvec_{1\Mvec}^\Delta[\Mvec_1]$ is the modification of the
free-space magnetostatic field $\Hvec_\Mvec[\Mvec_1]$ that accounts for the
shielding effect of the cavity walls.
It corresponds to the perfect-conductor limit of the wall effect
observed in ferrimagnetic resonance experiments \cite{NashStandley1960} and
analyzed for walls of finite conductivity by the method of images
\cite{Srivastava1966}.

\section{Lagrangian and Hamiltonian Field Formulation}
\label{sec:L_and_H_of_the_coupled_problem}

\subsection{Electric vector potential}
\label{sec:electric_C}
The Lagrangian formulation of the electromagnetic field interacting with
matter is conventionally expressed in terms of the magnetic vector potential
and the electric scalar potential in the Coulomb gauge
\cite{Jackson1999,CohenTannoudji1989}. In this formulation, the independent
electromagnetic degrees of freedom are associated with the transverse field,
while the longitudinal electric field is determined by the charge density.
In the present problem, where the charge density vanishes, the
magnetization-generated field $\Bvec_\Mvec^V[\Mvec_1]$ is itself solenoidal
and therefore contributes to the transverse field represented by the magnetic
vector potential. In particular,
\begin{align}
	\Bvec_1
	=
	\mu_0\Hvec_\text{1R}
	+
	\Bvec_\Mvec^V[\Mvec_1]
	=
	\nabla\times\Avec_1,
\end{align}
with the magnetization acting as a source through its equivalent Amp\`erian
current density
$\Jvec_{\Mvec_1}=\nabla\times\Mvec_1$ \cite{Bobbio1999}
(see Appendix~\ref{Sec:vector_pot_form_polaritons}).

The linearized LL equation, on the other hand, depends on the combination
\begin{align}
	\nabla\times\Avec_1-\Bvec_\Mvec[\Mvec_1]
	=
	\mu_0\left(
	\Hvec_\text{1R}
	+
	\Hvec^\Delta_{1\Mvec}[\Mvec_1]
	\right),
\end{align}
where
$\Bvec_\Mvec[\Mvec_1]
=\mu_0\left(\Hvec_\Mvec[\Mvec_1]+\Mvec_1\right)$
is the free-space field generated by the magnetization. Thus, in a truncated
cavity-mode representation, the physically relevant field is obtained through
the cancellation of the magnetization-generated contribution contained in
$\nabla\times\Avec_1$, namely $\Bvec_\Mvec^V[\Mvec_1]$, against its
free-space counterpart $\Bvec_\Mvec[\Mvec_1]$. Since this localized field converges slowly in a
cavity-mode expansion, the cancellation is poorly reproduced under modal
truncation, and the resulting error can greatly exceed the comparatively
small cavity-induced correction
(see Appendix~\ref{Sec:vector_pot_form_polaritons}).

The formulation adopted here is the dual of the Coulomb-gauge $\Avec$-formulation.  Since, in
eqs.~\eqref{eq:MaxwellrotE1_EB_coup_bR}--\eqref{eq:MaxwelldivH1_EB_coup_bR},
the Maxwell--Amp\`ere equation is homogeneous and the source term appears
only in Faraday's law, this field is naturally described by an electric
vector potential $\Cvec_1$, defined through
\begin{align}
	\label{eq:def_of_C1}
	\nabla \times \Cvec_1 = -\varepsilon_0\Evec_1 \, , \qquad
	\nabla \cdot \Cvec_1 = 0
	\qquad \text{in } V \, ,
\end{align}
together with the boundary condition
\begin{equation}
	\label{eq:BC_of_C1_1}
	\nvec \cdot \Cvec_1 = 0
	\qquad \text{on } \partial V \, ,
\end{equation}
which constitute the electric counterpart of the Coulomb gauge. The quantity
$\Cvec_1$ is known as the FitzGerald electric vector potential in classical
electromagnetic theory \cite{Someda2006}. Electric vector potentials also
appear in quantum electrodynamics with magnetic monopoles
\cite{Zwanziger1971,BlagojevicSenjanovic1988} and in dual-symmetric
formulations of classical electromagnetism \cite{Bliokh2013}.

The conditions~\eqref{eq:def_of_C1} and~\eqref{eq:BC_of_C1_1} uniquely determine the vector field $\Cvec_1$ for any admissible electric field $\Evec_1$.
Furthermore, the boundary condition~\eqref{eq:E_boundary_cond_E1} implies
\begin{align}
	\label{eq:BC_of_C1_2}
	\nvec \times \nabla \times \Cvec_1 = 0
	\qquad \text{on } \partial V \, .
\end{align}
Substituting the first of eqs.~\eqref{eq:def_of_C1} into
eq.~\eqref{eq:MaxwellrotB1_EB_coup_bR}, we obtain
\begin{align}
	\label{eq:B1_in_terms_C1}
	\Hvec_\text{1R} =  -\partial_t \Cvec_1 \, .
\end{align}
The relations~\eqref{eq:def_of_C1}--\eqref{eq:B1_in_terms_C1}
automatically satisfy
eqs.~\eqref{eq:MaxwellrotB1_EB_coup_bR},
\eqref{eq:MaxwelldivE1_EB_coup_bR},
\eqref{eq:MaxwelldivH1_EB_coup_bR},
together with the boundary conditions
\eqref{eq:E_boundary_cond_E1} and
\eqref{eq:E_boundary_cond_B1R}.

Accordingly, the vector potential $\Cvec_1$ belongs to the linear space
\begin{align}
	\nonumber
	&\mathbb{H}^{\mathrm{em}}(V)
	= \big\{ \uvec(\xvec) \, : \,
	\nabla \cdot \uvec = 0 \,\, \text{in } V \,, \,
	\nvec \cdot \uvec = 0 \,\, \text{on } \partial V \,, \\[2pt]
	\label{eq:vector_potential_space}
	& \quad
	\nvec \times \nabla \times \uvec = 0 
	\,\, \text{on } \partial V \,, \,
	\{\uvec,\nabla \times \uvec\} \in \mathbb{L}^2(V)
	\big\} \, ,
\end{align}
which consists of divergence-free vector fields in $V$ with square-integrable components and curls, satisfying a vanishing normal component and vanishing tangential curl on the boundary $\partial V$.
As will be discussed later in the manuscript, the space
$\mathbb{H}^{\mathrm{em}}(V)$ can be identified with the linear space of electromagnetic cavity modes describing the free oscillations of the cavity field.

By using eqs.~\eqref{eq:def_of_C1} and \eqref{eq:B1_in_terms_C1} in
eq.~\eqref{eq:MaxwellrotE1_EB_coup_b}, we obtain
\begin{align}
	\label{eq:A1evolution_1_coup_A}
	c^{-2} \partial_t^2 \Cvec_1
	-
	\nabla^2 \Cvec_1
	=
	\varepsilon_0 \partial_t \Bvec^V_\Mvec[\Mvec_1] \, ,
\end{align}
while the constitutive relation~\eqref{eq:general_VLL_Gqs_lin_bis} becomes
\begin{align}
	\label{eq:general_VLL_Gqs_lin_c}
	\partial_t \Mvec_1
	=
	\gamma \Mvec_0 \times
	\left(
	\delta_{\Mvec_1} \mathcal{H}_\text{M}
	+
	\mu_0 \partial_t \Cvec_1
	-
	\mu_0 \Hvec_{1\Mvec}^\Delta
	\right) .
\end{align}
Equations~\eqref{eq:A1evolution_1_coup_A} and
\eqref{eq:general_VLL_Gqs_lin_c} therefore define the coupled linear
evolution problem in terms of the dynamical variables
$\Cvec_1$, $\partial_t\Cvec_1$, and $\Mvec_1$, subject to the constraints
$\Cvec_1\in\mathbb{H}^{\mathrm{em}}(V)$
(see eq.~\eqref{eq:vector_potential_space}) and
$\Mvec_1\in\mathbb{H}^{\mathrm{sw}}(V_\text{M})$
(see eq.~\eqref{eq:spin-wave_space_f}).
In this formulation, only $\Cvec_1$ needs to be represented in terms of
cavity modes, and the cancellation problem discussed above is avoided.

\subsection{Field Lagrangian}

\label{sec:field_L}

The coupled evolution equations
\eqref{eq:A1evolution_1_coup_A} and \eqref{eq:general_VLL_Gqs_lin_c}
follow from the Lagrangian
\begin{widetext}
	\begin{align}
		\nonumber
		L= &
		\int_V \!\left[\frac{\mu_0 }{2} \left(\partial_t \Cvec_1\right)^2 - \frac{1}{2\varepsilon_0} (\nabla \times \Cvec_1)^2
		\right] dV + \int_{V_\text{M}} \!\! \frac{1}{2\gamma\Msat^2}
		\left\{
		\Mvec_0 \times \Mvec_1\cdot \partial_t \Mvec_1
		\right\} dV
		- \mathcal{H}_\text{M}[\Mvec_1] \\[3pt]
		& \qquad \qquad \qquad \qquad\qquad \qquad \quad
		+ \int_{V_\text{M}} \frac{\mu_0}{2} \Mvec_1 \cdot \Hvec_{1\Mvec}^\Delta[\Mvec_1]  \, dV
		+ \int_V \Cvec_1\cdot  \partial_t \Bvec_\Mvec^V[\Mvec_1] \, dV
		\, ,
		\label{eq:cavity_fild_lagrangian_coup_0_f}
	\end{align}
\end{widetext}
which naturally decomposes into three contributions: the electromagnetic
field part $L_\text{P}$, given by the first integral, the magnetic part
$L_\text{M}$, given by the gyrotropic term together with
$-\mathcal{H}_\text{M}[\Mvec_1]$, and the interaction part
$L_\text{int}$, consisting of the last two terms. These two interaction
contributions will be denoted by $L^\Delta_\text{int}$ and
$L^Z_\text{int}$, respectively, in the following.
The magnetic part is the linearization around $\Mvec_0$
of the standard Lagrangian of magnetization dynamics
\cite{Gilbert2004,Ruijgrok1980,Valet2026}, whose kinetic term is first order
in the time derivative of the magnetization.

The last term in eq.~\eqref{eq:cavity_fild_lagrangian_coup_0_f}
describes the interaction of the magnetization with the cavity field.
Up to a total time derivative, which does not affect the Euler--Lagrange equations, it can be rewritten as
\begin{align}
	\label{eq:L_int_1st_0}
	-\int_V \Bvec_\Mvec^V \cdot \partial_t \Cvec_1 \, dV
	=
	\mu_0 \int_V \Mvec_1 \cdot \Hvec_\text{1R} \, dV \, ,
\end{align}
which has the familiar form of the Zeeman interaction between the magnetization and the radiative component of the magnetic field. Equation~\eqref{eq:L_int_1st_0} follows from the Helmholtz decomposition~\eqref{eq:Helmholtz_decomp_M1_0}  together with the identity~\eqref{eq:Tellegen_th}.

The term $L^\Delta_\text{int}$ associated with the field $\Hvec_\Mvec^\Delta$ accounts for the
modification of the magnetostatic field induced by the shielding effect of
the cavity walls, discussed at the end of Sec.~\ref{sec:Decomposition_of_B1}.
The corresponding additional magnetostatic energy is positive
semidefinite, as shown in Appendix~\ref{Sec:positive_H_delta_M}.

Since the interaction mediated by $\Hvec_\Mvec^\Delta$ depends only on the
magnetization, its contribution could, in principle, be regarded as part of
the magnetic Lagrangian. The distinction is ultimately a matter of how the free
magnetic subsystem is defined. In the present work, we take the free magnetic
dynamics to be that of the magnetic body isolated in free space and treat the
cavity-induced magnetostatic correction as an interaction term. This choice
has the practical advantage that the free magnetic modes can be computed
directly using standard micromagnetic methods
\cite{DonahuePorter1999,Vansteenkiste2014MuMax3,Abert2013MagnumFE,Fischbacher2007Nmag,Schrefl2007}.

The Euler--Lagrange equation associated with $\Cvec_1$, obtained from the Lagrangian~\eqref{eq:cavity_fild_lagrangian_coup_0_f} by restricting the admissible variations to $\Cvec_1 \in \mathbb{H}^\text{em}(V)$ and $\Mvec_1 \in \mathbb{H}^\text{sw}(V_\text{M})$, is
\begin{align}
	\mu_0 \varepsilon_0 \partial_t^2 \Cvec_1
	+ \nabla \times \nabla \times \Cvec_1
	- \varepsilon_0\partial_t \Bvec_\Mvec^V
	= 0 \, ,
\end{align}
which, because $\nabla\cdot\Cvec_1=0$, is equivalent to Eq.~\eqref{eq:A1evolution_1_coup_A}.

The Euler--Lagrange equation associated with $\Mvec_1$ takes the form
\begin{align}
	\nonumber
	\frac{1}{\gamma\Msat^2}
	\Mvec_0 \times \partial_t \Mvec_1
	&=
	-\delta_{\Mvec_1}\mathcal{H}_\text{M} \\
	&\quad
	- \mu_0\partial_t \Cvec_{1\bot}
	+ \mu_0\Hvec_{1\Mvec\bot}^{\Delta} \, ,
	\label{eq:EL_eqs_M1}
\end{align}
where
$\Cvec_{1\bot}=\mathbbm{P}_{0\bot}(\xvec)\cdot\Cvec_1$
and
$\Hvec_{1\Mvec\bot}^{\Delta}
=\mathbbm{P}_{0\bot}(\xvec)\cdot\Hvec_{1\Mvec}^{\Delta}$,
with $\mathbbm{P}_{0\bot}(\xvec)$ defined in
Eq.~\eqref{eq:def_of_D0bot_f_1}. The projector $\mathbbm{P}_{0\bot}(\xvec)$ enforces the constraint
$\Mvec_1 \cdot \Mvec_0=0$.

In deriving Eq.~\eqref{eq:EL_eqs_M1}, we have used the reciprocity properties of the fields
$\Bvec_\Mvec[\Mvec_1]=\mu_0(\Hvec_\Mvec[\Mvec_1]+\Mvec_1)$ \cite{Brown1962} and
$\Bvec_\Mvec^V[\Mvec_1]$ (see Eq.~\eqref{eq:reciprocity_in_V}),  together with the identity
\begin{align}
	\label{eq:identity_int_term_CM}
L^Z_\text{int}=	\int_V \Cvec_1\cdot  \partial_t \Bvec_\Mvec^V \, dV
	=  \mu_0 \int_V \Cvec_1\cdot  \partial_t \Mvec_1 \, dV \, ,
\end{align}
which follows from arguments analogous to those used in the derivation of Eq.~\eqref{eq:L_int_1st_0}.

Finally, using the orthogonality condition
$\Mvec_1(\xvec,t) \cdot \Mvec_0(\xvec)=0$,
eq.~\eqref{eq:EL_eqs_M1} reduces to
eq.~\eqref{eq:general_VLL_Gqs_lin_c}
upon taking the vector product of both sides with $\Mvec_0$.

\subsection{Field Hamiltonian}
\label{sec:field_H}

As a first step, we compute the functional derivatives of the Lagrangian
with respect to the time derivatives of the fields:
\begin{align}
	\label{eq:conj_momentum_A1}
	\bm{\Pi}_{\Cvec_1}
	&=
	\frac{\delta L}{\delta (\partial_t \Cvec_1)}
	=
	\mu_0 \partial_t \Cvec_1 \, , \\[3pt]
	\label{eq:cnj_mom_M1}
	\bm{\Pi}_{\Mvec_1}
	&=
	\frac{\delta L}{\delta (\partial_t \Mvec_1)}
	=
	\frac{1}{2\gamma M_{\mathrm{s}}^{2}}
	\Mvec_0 \times \Mvec_1
	+\mu_0 \Cvec_1,
\end{align}
where, in deriving Eq.~\eqref{eq:cnj_mom_M1}, we have used the identity
\eqref{eq:identity_int_term_CM}.

Since the Lagrangian depends linearly on $\partial_t\Mvec_1$, the canonical
momentum density $\bm{\Pi}_{\Mvec_1}$ does not depend on
$\partial_t\Mvec_1$. Consequently, the Legendre transformation is singular
in the magnetic sector: the relation defining $\bm{\Pi}_{\Mvec_1}$ cannot
be inverted to express $\partial_t\Mvec_1$ in terms of the canonical
momentum. Nevertheless, the Hamiltonian can be introduced as
\begin{align}
	\mathcal{H}
	=
	\int_V
	\left( \bm{\Pi}_{\Cvec_1}\cdot\partial_t\Cvec_1\,dV
	+
	\bm{\Pi}_{\Mvec_1}\cdot\partial_t\Mvec_1\right) \,dV
	-L \, .
	\label{eq:Hamiltonian_C_pi_C}
\end{align}
The terms proportional to $\partial_t\Mvec_1$ cancel as a consequence of
the first-order structure of the magnetic Lagrangian, leading to
\begin{align}
	\mathcal{H}
	=
	\int_V
	\left[
	\frac{(\bm{\Pi}_{\Cvec_1})^2}{2\mu_0}
	+
	\frac{(\nabla\times\Cvec_1)^2}{2\varepsilon_0}
	\right]
	dV
	+
	\widetilde{\mathcal{H}}_{\mathrm{M}},
	\label{eq:Hamiltonian_lin_coupled_0_f_a}
\end{align}
where
\begin{align}
	\widetilde{\mathcal{H}}_{\mathrm{M}}
	=
	\mathcal{H}_{\mathrm{M}}
	+
	\mathcal{H}_{\mathrm{M}}^\Delta \, ,
	\label{eq:Hamiltonian_lin_coupled_0_f_a_bis}
\end{align}
and
\begin{align}
	\mathcal{H}_{\mathrm{M}}^\Delta
	=
	-
	\int_{V_{\mathrm{M}}}
	\frac{\mu_0}{2}
	\Mvec_1\cdot
	\Hvec_{1\Mvec}^{\Delta}[\Mvec_1]
	\,dV \, .
	\label{eq:Hamiltonian_lin_coupled_0_f_delta}
\end{align}
Equation~\eqref{eq:Hamiltonian_lin_coupled_0_f_a} naturally separates into
an electromagnetic contribution, given by the volume integral over $V$, and
the magnetic contribution $\widetilde{\mathcal{H}}_{\mathrm{M}}$. The latter
consists of the free-space micromagnetic Hamiltonian
$\mathcal{H}_{\mathrm{M}}$ together with the additional positive
semidefinite magnetostatic contribution
$\mathcal{H}_{\mathrm{M}}^\Delta$
(see Appendix~\ref{Sec:positive_H_delta_M}), arising from the modification
(shielding) of the dipolar field by the cavity walls.

 The interaction between the radiative field $\Hvec_\text{1R}$ and the magnetization does not appear explicitly in the Hamiltonian.
Instead, this interaction is encoded in the noncanonical structure of the Hamiltonian system. To avoid technical complications associated with functional derivatives, this aspect will be discussed in Sec.~\ref{sec:finite_dim_NM_L_and_H}, after introducing the normal-mode description.

Despite these unconventional features, it is important to emphasize that $\mathcal{H}$ retains the meaning of the total energy of the system and, for this reason, is an integral of motion. By using the expressions \eqref{eq:operator_D1_f_energy0} and \eqref{eq:operator_D1_f_energy} for $\mathcal{H}_\text{M}[\Mvec_1]$, together with the definition \eqref{eq:def_of_H1delta} of $\Hvec_{1\Mvec}^\Delta[\Mvec_1]$, the Hamiltonian can be recast in the form
\begin{align}
	\label{eq:energy_H}
	\mathcal{H} = F_0[\Mvec_1] + \int_V \left(\frac{\mu_0}{2}\Hvec_1^2 + \frac{\varepsilon_0}{2}\Evec_1^2\right)dV  \, ,
\end{align}
where we have used the relations $\nabla \times \Cvec_1 = -\varepsilon_0\Evec_1$,
$-\partial_t\Cvec_1 + \Hvec_\Mvec^V[\Mvec_1]= \Hvec_1$ and the fact that $\int_V {\bm \Pi}_{\Cvec_1} \cdot \Hvec_M^V[\Mvec_1]dV =0$.
One can then readily verify that, by using Poynting's theorem \cite{FanoChuAdler1960} for the system
\eqref{eq:MaxwellrotE1_EB_coup_bR}--\eqref{eq:MaxwelldivH1_EB_coup_bR}, with boundary condition \eqref{eq:E_boundary_cond_E1}, together with the constitutive relation
\eqref{eq:general_VLL_Gqs_lin_c}, the expression on the right-hand side of eq.~\eqref{eq:energy_H}
has vanishing time derivative.

We conclude by noting that the Lagrangian term describing the interaction with the
radiative field could equivalently be written in the form~\eqref{eq:L_int_1st_0}.
This alternative choice would lead to the same equations of motion for the
coupled system. However, it would imply the relation
${\bm \Pi}_{\Cvec_1}=-\Bvec_1$
(see Appendix~\ref{sec:C_formulation_polartiton}). The corresponding
Hamiltonian formulation would then lead to an evolution equation for
$\Mvec_1$ in which the cancellation issues discussed at the beginning
of Sec.~\ref{sec:electric_C} reappear.

\section{Normal modes of noninteracting subsystems}

\label{sec:normal_modes_non_int}

\subsection{Normal modes of the free magnon dynamics}

As discussed in Sec.~\ref{sec:field_L}, we define the free magnetization
dynamics as that of the magnetic body in the absence of the cavity, with
$\Hvec_1=\Hvec_\Mvec[\Mvec_1]$ (see Sec.~\ref{sec:Decomposition_of_B1}). The
corresponding magnetic Lagrangian $L_\text{M}$ is obtained from
eq.~\eqref{eq:cavity_fild_lagrangian_coup_0_f} by setting
$\Cvec_1=0$ and $\Hvec_{1\Mvec}^{\Delta}=0$.
The resulting dynamics of $\Mvec_1$ is governed by the linearized
Landau--Lifshitz equation
\begin{align}
	\label{eq:Lin_LL_equation_0c_f}
	\frac{\partial \Mvec_1}{\partial t} =
	\gamma \mu_0 \Msat \, \mvec_0(\xvec) \times
	{\mathbbm{D}}_{0\bot}[\Mvec_1] \, ,
\end{align}
where $\mathbbm{D}_{0\bot}[\,\cdot\,]$ is defined by
eqs.~\eqref{eq:operator_D1_f} and~\eqref{eq:def_of_D0bot_f}.
As discussed in Sec.~\ref{sec:linear_mag_dyn},
the operator $\mathbbm{D}_{0\bot}[\,\cdot\,]$ is Hermitian and
positive definite.

 Equation~\eqref{eq:Lin_LL_equation_0c_f} 
can be written more compactly as
\begin{align}
	\label{eq:Linearized_LL_from_L_1_f}
	\frac{\partial \Mvec_1}{\partial t} =  \omega_s \, {\bm \Lambda}_0 \,{\mathbbm{D}}_{0\bot}[\Mvec_1] \, ,
\end{align}
where $\omega_s = \gamma \mu_0 \Msat$, and where ${\bm \Lambda}_0$ denotes 
the tensor representing the pointwise vector product with $\mvec_0(\xvec)$. 
In eqs.~\eqref{eq:Lin_LL_equation_0c_f}--\eqref{eq:Linearized_LL_from_L_1_f},
the vector field $\Mvec_1$ belongs to the spin-wave space $\mathbb{H}^\text{sw}(V_\text{M})$
(see eq.~\eqref{eq:spin-wave_space_f}), and the operator $\mathbbm{D}_{0\bot}[\,\cdot\,]$ 
(see eqs.~\eqref{eq:operator_D1_f}, \eqref{eq:def_of_D0bot_f}) is, as discussed in Sec.~\ref{sec:linear_mag_dyn},
positive definite, real, and self-adjoint.

To determine the normal modes, we seek solutions of~\eqref{eq:Linearized_LL_from_L_1_f} of the form
\begin{align}
	\label{eq:complex_amplitude_repr_0_f}
	\Mvec_{1}(\xvec,t) 
	=\bar{M}_m \, e^{-i\omega_m t} \, \varphivec_m(\xvec) \, ,
\end{align}
where $\omega_m$ is the angular frequency of the $m$-th magnetization normal mode,
$\varphivec_m(\xvec)$ is an element of the space $\mathbb{H}^\text{sw}(V_\text{M})$,
and $\bar{M}_m$ is a complex constant. 

Substituting~\eqref{eq:complex_amplitude_repr_0_f} into~\eqref{eq:Linearized_LL_from_L_1_f} yields the normal-mode equation 
\begin{align}
	\label{eq:llg_normal_mode_f}
	\nu_m \, \varphivec_m =  i{\bm \Lambda}_0 \, \mathbbm{D}_{0\bot}[\varphivec_m] \, ,
\end{align}
where $\nu_m = \omega_m/\omega_s$ are the normalized frequencies.
This is the eigenvalue problem for the non-Hermitian operator $i{\bm \Lambda}_0 \, \mathbbm{D}_{0\bot}$.

Multiplying both sides of~\eqref{eq:llg_normal_mode_f} by $i{\bm \Lambda}_0$, 
we can rewrite the eigenvalue equation in the equivalent form
\begin{equation}
	\label{eq:llg_normal_mode_0_f}
	\nu_m \, i {\bm \Lambda}_0 \, \varphivec_m = \mathbbm{D}_{0\bot}[\varphivec_m]  \, ,
\end{equation}
which represents a generalized eigenvalue problem involving two Hermitian operators: 
$\mathbbm{D}_{0\bot}[\,\cdot\,]$ and $i{\bm \Lambda}_0$. 
From the Hermitian nature of the two operators appearing in~\eqref{eq:llg_normal_mode_0_f},
and the fact that $\mathbbm{D}_{0\bot}[\,\cdot\,]$ is positive definite, 
one can show, using standard arguments, that the eigenvalues $\nu_m$ are real
\cite{dAquino_2009,HornJohnson2013}.

In addition, the eigensolutions of eq.~\eqref{eq:llg_normal_mode_0_f} are denumerable and occur in pairs: 
if $(\nu_m, \varphivec_m)$ is a solution of the eigenvalue problem, then $(-\nu_m, \varphivec_m^*)$ is also a solution.  

For convenience and to simplify notation, we introduce the following labeling convention:
\begin{align}
	\nu_{-m} = -\nu_{m}  \, , \quad 
	\varphivec_{-m}(\xvec) = \varphivec_{m}^*(\xvec) \, , 
	\quad m = 1, 2, \ldots
\end{align}
We then consider the sequence $(\nu_k, \varphivec_k(\xvec))$ with $k \in \mathbb{Z}$, 
where $k$ may take both positive and negative values. By convention, we set $\nu_0 = 0$ 
and $\varphivec_0(\xvec) = 0$, consistently with the assumption that $\mathbbm{D}_{0\bot}[\, \cdot\,]$ is
nonsingular. In addition, we assume that $\nu_k > 0$ for $k > 0$.

As a further consequence of the structure of the problem~\eqref{eq:llg_normal_mode_0_f}, the vector fields $\varphivec_k(\xvec)$
form a basis of the spin-wave space $\mathbb{H}^\text{sw}(V_\text{M})$ and satisfy the following
Walker-type generalized orthogonality conditions
\cite{Walker1957,Mills2006,SerpicoPRB2025,Valet2026}
\begin{align}
	\label{eq:Linearized_LL_modes_3a_f}
	& \frac{1}{|V_\text{M}|}\int_{V_\text{M}} \varphivec_h^*(\xvec)\cdot i{\bm \Lambda}_0  \, \varphivec_k(\xvec) \, dV =
	\text{sgn}(k)\delta_{kh} \,, \\
	\label{eq:Linearized_LL_modes_3b_f}
	& \frac{1}{|V_\text{M}|}\int_{V_\text{M}} \varphivec_h^*(\xvec)\cdot  \mathbbm{D}_{0\bot} [ \varphivec_k(\xvec)] \, dV =
	\nu_{|k|} \delta_{kh} \, ,
\end{align}
where $k,h \in \mathbb{Z}$, and $|V_\text{M}|$ is the volume of $V_\text{M}$.

Since the set of eigenfunctions is also complete, we can expand the vector field $\Mvec_1$ in the
bilateral generalized Fourier series
\begin{align}
	\label{eq:Fourier_series_bilat_M_0_f}
	\Mvec_1(\xvec,t) = \sum_{k\in \mathbb{Z}} M_k(t) \varphivec_k(\xvec) \, ,
\end{align}
where 
\begin{align}
	\label{eq:Fourier_coeff_bilat_M_1_f}
	M_k(t) = \frac{\text{sgn}(k)}{|V_\text{M}|} \int_{V_\text{M}} \varphivec_k^*(\xvec)\cdot  i{\bm \Lambda}_0 \Mvec_1(\xvec,t) \, dV \, ,
\end{align}
with $k \in \mathbb{Z}$.

Substituting the expansion~\eqref{eq:Fourier_series_bilat_M_0_f} into the
magnetic part of the Lagrangian, namely the third and fourth terms in
eq.~\eqref{eq:cavity_fild_lagrangian_coup_0_f}, and using the orthogonality
relations~\eqref{eq:Linearized_LL_modes_3a_f}--\eqref{eq:Linearized_LL_modes_3b_f},
we obtain the normal-mode Lagrangian
\begin{align}
	\nonumber
	L_\text{M} =& \sum_{m=1}^\infty \frac{i\hbar }{2}\left( b_m^* \dot{b}_m - b_m \dot{b}_m^* \right)  \\
	&-\sum_{m =1}^\infty
	\frac{\hbar \omega_{m}}{2}
	\left[ b_m(t) b^*_m(t) + b^*_m(t) b_m(t) \right] \, ,
	\label{eq:Lagrangian_bs}
\end{align}
where
\begin{align}
	\label{eq:bosonic_b_for_magnons}
	b_m(t)
	= \sqrt{\frac{|V_\text{M}|}{\gamma \hbar \Msat}} \, M_m(t) \, ,
	\qquad m \in \mathbb{N} \, ,
\end{align}
and $\omega_m = \gamma \mu_0 \Msat \nu_m$.

The corresponding Hamiltonian is obtained from the Lagrangian \eqref{eq:Lagrangian_bs} according to
\begin{align}
	\mathcal{H}_\text{M}
	=
	\sum_m\left(
	\frac{\partial L_\text{M}}{\partial \dot{b}_m} \dot{b}_m
	+
	\frac{\partial L_\text{M}}{\partial \dot{b}^*_m} \dot{b}_m^*
	\right)
	- L_\text{M} \, ,
\end{align}
and is given by
\begin{align}
	\label{eq:sw_energy_tris_0_3_f}
	\mathcal{H}_\text{M}
	= \sum_{m \in \mathbb{N}}
	\frac{\hbar \omega_{m}}{2}
	\left[ b_m(t) b^*_m(t) + b^*_m(t) b_m(t) \right] \, .
\end{align}

The Euler--Lagrange equations associated with the Lagrangian \eqref{eq:Lagrangian_bs}
are
\begin{align}
	\label{eq:eqs_lin_dyn_classical_bosonic_0_f}
	& \dot b_m = - i \omega_m b_m = \frac{1}{i\hbar} \frac{\partial \mathcal{H}_\text{M}}{\partial b_m^*} \, , \\
	& \dot{b}^*_m =  i \omega_m b^*_m = -\frac{1}{i\hbar} \frac{\partial \mathcal{H}_\text{M}}{\partial b_m} \, ,
\end{align}
with $m \in \mathbb{N}$. These equations can be recast in Hamiltonian form as
\begin{align}
	\dot b_m = \{b_m,\mathcal{H}_\text{M}\} \, , \qquad
	\dot{b}^*_m = \{b_m^*,\mathcal{H}_\text{M}\} \, ,
\end{align}
where $\{ \cdot\, , \cdot\}$ denotes the Poisson bracket associated with the Lagrangian
\eqref{eq:Lagrangian_bs} \cite{Faddeev1988}, namely
\begin{align}
	\label{eq:PBs_bosonic_M_f}
	\{f,g\}
	= \frac{1}{i\hbar}
	\sum_{m \in \mathbb{N}}
	\left(
	\frac{\partial f}{\partial b_m}
	\frac{\partial g}{\partial b^*_m}
	-
	\frac{\partial f}{\partial b^*_m}
	\frac{\partial g}{\partial b_m}
	\right) ,
\end{align}
where the variables $(b_m,b_m^*)$ are treated as independent when evaluating the derivatives.

Using~\eqref{eq:PBs_bosonic_M_f}, we obtain the canonical Poisson brackets
\begin{align}
	\label{eq:bosonic_PB_bs}
	\{b_m,b^*_h\} = \frac{\delta_{mh}}{i\hbar} \, , \quad
	\{b_m,b_h\} = \{b_m^*,b_h^*\} = 0 \, ,
\end{align}
with $m,h \in \mathbb{N}$.

Equation~\eqref{eq:sw_energy_tris_0_3_f} shows that the Hamiltonian corresponds to a collection of independent harmonic oscillators,
and the Poisson brackets \eqref{eq:bosonic_PB_bs} show that $(b_m,b_m^*)$ are the classical analogues of bosonic annihilation and creation operators for magnons. In this sense, the Lagrangian \eqref{eq:Lagrangian_bs} is the classical counterpart of the bosonic Lagrangian describing a free field.

Finally, the magnetization field can be expressed in terms of the modal variables as
\begin{align}
	\nonumber
	\Mvec_1(\xvec,t) 
	= \sum_{m\in \mathbb{N}} 
	\sqrt{\frac{\gamma \hbar \Msat}{|V_\text{M}|}}
	\bigl[ & b_m(t) \varphivec_m(\xvec) \\ &+ b_m^*(t) \varphivec_m^*(\xvec)\bigr] \, ,
	\label{eq:norma_modes_M1_expansion}
\end{align}
where we have used eqs.~\eqref{eq:Fourier_series_bilat_M_0_f} and
\eqref{eq:bosonic_b_for_magnons}.

\subsection{Normal modes of free cavity electromagnetic fields}

We consider the natural oscillations of the cavity electromagnetic field
described by eq.~\eqref{eq:A1evolution_1_coup_A} in the absence of a magnetic body, i.e.,
for $\Mvec_1 = 0$. In this case, using the condition $\nabla \cdot \Cvec_1 = 0$,
the evolution equation for $\Cvec_1$ can be rewritten as
\begin{align}
	\label{eq:C1evolution_1_free_rot_f}
	c^{-2}\, \partial_t^2 \Cvec_1  + \nabla \times \nabla \times \Cvec_1 = 0
	\qquad \text{in } V \, ,
\end{align}
with boundary conditions \eqref{eq:BC_of_C1_1}, \eqref{eq:BC_of_C1_2}.
Notice that, for any solution of eq.~\eqref{eq:C1evolution_1_free_rot_f}, the boundary condition
\eqref{eq:BC_of_C1_1} is redundant, since it can be derived from condition \eqref{eq:BC_of_C1_2} together with
eq.~\eqref{eq:C1evolution_1_free_rot_f}.

To determine the normal modes, we consider the Ansatz
\begin{align}
	\label{eq:ansatz_free_oscillations_f}
	\Cvec_{1}(\xvec,t) =
	\bar{C}_p e^{-i \widetilde{\omega}_p t} \hvec_p(\xvec) \, ,
\end{align}
where $p$ is a collective index labeling the cavity modes,
$\widetilde{\omega}_p$ is the corresponding angular frequency,
$\hvec_p(\xvec)$ is an element of $\mathbb{H}^{\mathrm{em}}(V)$, and
$\bar{C}_p$ is a complex constant.

Substituting eq.~\eqref{eq:ansatz_free_oscillations_f} into eq.~\eqref{eq:C1evolution_1_free_rot_f},
we obtain the eigenvalue problem for magnetic solenoidal modes:
\begin{align}
	\label{eq:electric_sol_modes_V_0_f}
	& \nabla \times \nabla \times \hvec_p(\xvec) = k_p^2 \hvec_p(\xvec) && \text{in } V \, ,\\
	\label{eq:electric_sol_modes_S_0_f}
	& \nvec \times \nabla \times \hvec_p = 0 && \text{on } \partial V \, ,
\end{align}
where $k_p = \widetilde{\omega}_p / c$ is real and positive \cite{Kurokawa1958}.

Once the discrete eigenvalues $k_p$ have been determined, the corresponding frequencies
$\widetilde{\omega}_p = c k_p$ represent the allowed oscillation frequencies of the cavity fields.

The eigenfunctions $\hvec_p(\xvec)$ can be chosen to be real and form a complete orthonormal set, normalized as
\begin{align}
	\nonumber
	&	\frac{1}{|V|} \int_V \hvec_p(\xvec) \cdot \hvec_{p'}(\xvec) \, dV
	= \\
	& \quad \frac{1}{|V| k_p^2} \int_V \!\left[ \nabla \times \hvec_p(\xvec) \right] \!
	\cdot \!\left[ \nabla \times \hvec_{p'}(\xvec) \right] dV
	= \delta_{pp'} \, ,
	\label{eq:orthogonality_relations}
\end{align}
for $p,p' = 1,2,\ldots$ \cite{Kurokawa1958}, and $|V|$ is the volume of $V$. This set forms a complete basis for the subspace of transverse fields in the cavity.
Hence,
the vector potential $\Cvec_1$ can be expanded as
\begin{align}
	\label{eq:expression_of_C_in_terms_of_electric_modes_f}
	\Cvec_1(\xvec,t) = \sum_{p=1}^\infty C_p(t) \, \hvec_p(\xvec) \, ,
\end{align}
where $C_p(t)$ are real-valued functions. Moreover,
since $\hvec_p(\xvec) \in \mathbb{H}^{\mathrm{em}}(V)$, the field $\Cvec_1(\xvec,t)$ defined in
eq.~\eqref{eq:expression_of_C_in_terms_of_electric_modes_f} also belongs to $\mathbb{H}^{\mathrm{em}}(V)$.
Substituting the expansion~\eqref{eq:expression_of_C_in_terms_of_electric_modes_f} into the free cavity-field Lagrangian, given by the first two terms of the total Lagrangian \eqref{eq:cavity_fild_lagrangian_coup_0_f}, and using the orthogonality relations \eqref{eq:orthogonality_relations}, yields
\begin{align}
	\label{eq:mode_lagrangian_1_f}
	L_\text{P} = \sum_{p=1}^\infty
	\left[
	\frac{\hbar}{2\widetilde{\omega}_p}\dot q_p^2
	-\frac{\hbar\widetilde{\omega}_p}{2}q_p^2
	\right] \, ,
\end{align}
where
\begin{align}
	\label{eq:photon_mode_normalization_f}
	q_p=
	-\sqrt{\frac{\mu_0|V|\widetilde{\omega}_p}{\hbar}}\,
	C_p \, ,
\end{align}
and the dot denotes a time derivative.
With this normalization, $q_p$ is dimensionless.

The momenta conjugate to the variables $q_p$ are
\begin{align}
	\label{eq:photon_mode_p_normalization_f}
	p_p
	=
	\frac{\partial L_\text{P}}{\partial \dot{q}_p}
	=
	\frac{\hbar}{\widetilde{\omega}_p}\dot{q}_p
	=
	-\sqrt{\frac{\mu_0 |V|\hbar}{\widetilde{\omega}_p}} \, \dot{C}_p \, ,
\end{align}
and have dimension of $\hbar$.

The Hamiltonian expressed in terms of the modal variables is obtained from the Lagrangian \eqref{eq:mode_lagrangian_1_f} by means of the Legendre transformation \cite{Gantmacher1975} and is given by
\begin{align}
	\label{eq:Hamiltonian_of_modes_f}
	\mathcal{H}_\text{P} = \sum_{p=1}^\infty
	\left[
	\frac{\widetilde{\omega}_p}{2\hbar}p_p^2
	+ \frac{1}{2}\hbar\widetilde{\omega}_p q_p^2
	\right] \, ,
\end{align}
which corresponds to a set of independent harmonic oscillators with angular frequencies $\widetilde{\omega}_p$.

The Poisson brackets associated with the Lagrangian \eqref{eq:mode_lagrangian_1_f}
are canonical:
\begin{align}
	\{ f, g \} = \sum_{p=1}^\infty
	\left[
	\frac{\partial f}{\partial q_p} \frac{\partial g}{\partial p_p}
	- \frac{\partial f}{\partial p_p} \frac{\partial g}{\partial q_p}
	\right] \, ,
\end{align}
with $f$ and $g$ arbitrary functions of $( q_p, p_p )$.

Next, we reformulate the dynamics in terms of the classical analogues of
the creation and annihilation operators:
\begin{align}
	\label{eq:an_of_A1}
	a_p &= \frac{1}{\sqrt{2}}
	\left( q_p + i \frac{p_p}{\hbar} \right) \, , \\
	\label{eq:an*_of_A1}
	a_p^* &= \frac{1}{\sqrt{2}}
	\left( q_p - i \frac{p_p}{\hbar} \right) \, .
\end{align}

In terms of these variables, the Lagrangian can be written as
\begin{align}
	\nonumber
	\widetilde{L}_p
	=&
	\sum_{p=1}^\infty
	\frac{i\hbar}{2}
	\left(
	a_p^* \dot{a}_p
	-
	a_p \dot{a}_p^*
	\right)
	\\
	&
	-
	\sum_{p =1}^\infty
	\frac{\hbar \widetilde{\omega}_{p}}{2}
	\left[
	a_p(t) a^*_p(t)
	+
	a^*_p(t) a_p(t)
	\right] \, ,
	\label{eq:Lagrange_ph_as}
\end{align}
which is equal, up to a total time derivative, to the Lagrangian
\eqref{eq:mode_lagrangian_1_f}, as can be verified using
eqs.~\eqref{eq:an_of_A1} and \eqref{eq:an*_of_A1}.
This form is particularly convenient because it casts the cavity-mode
Lagrangian in the same symmetric first-order form as the magnetic-mode
Lagrangian~\eqref{eq:Lagrangian_bs}. This common structure allows the
Faddeev--Jackiw formalism to be applied in a unified manner to the magnetic
and electromagnetic modes when treating their hybridization in the next
section.

By proceeding in complete analogy with the magnetic normal modes discussed after
eq.~\eqref{eq:Lagrangian_bs}, we obtain the following expression for the cavity-field Hamiltonian:
\begin{align}
	\label{eq:Hamiltonian_modes_a_sym_f}
	\mathcal{H}_\text{P}
	=
	\sum_{p=1}^\infty
	\frac{\hbar \widetilde{\omega}_p}{2}
	\left( a_p a_p^* + a_p^* a_p \right) \, .
\end{align}
This expression can also be obtained by substituting the
expansion \eqref{eq:expression_of_C_in_terms_of_electric_modes_f}
into the free cavity-field Hamiltonian, given by
the first two terms on the right-hand side of eq.~\eqref{eq:Hamiltonian_lin_coupled_0_f_a},
and using eqs.\,\eqref{eq:conj_momentum_A1}, \eqref{eq:orthogonality_relations},
\eqref{eq:photon_mode_normalization_f}, \eqref{eq:an_of_A1},
\eqref{eq:an*_of_A1}.

From this point onward, equations analogous to
eqs.~\eqref{eq:eqs_lin_dyn_classical_bosonic_0_f}--\eqref{eq:bosonic_PB_bs}
can be obtained by replacing the variables $(b_m,b_m^*)$ with
$(a_p,a_p^*)$. For this reason, they will not be reported explicitly.

Finally, using eqs.~\eqref{eq:photon_mode_normalization_f},
\eqref{eq:photon_mode_p_normalization_f},
\eqref{eq:an_of_A1}, and \eqref{eq:an*_of_A1},
the vector potential $\Cvec_1(\xvec,t)$ can be expressed in terms of the complex
variables $a_p(t)$ and $a_p^*(t)$ as
\begin{align}
	\nonumber
	\Cvec_1(\xvec,t)
	= &-\sum_{p=1}^\infty
	\sqrt{
		\frac{\hbar}
		{2\mu_0 |V|\,\widetilde{\omega}_p}
	} \, \cdot \, \\
	&\qquad\qquad \cdot
	\left[ a_p(t) + a_p^*(t) \right] \hvec_p(\xvec) \, ,
	\label{eq:expression_of_A_in_terms_of_electric_modes_a_a^*_f}
\end{align}
where
\begin{align}
	a_p(t)=\bar a_p e^{-i\widetilde{\omega}_p t},
\end{align}
and $\bar a_p$ are complex constants determined by the initial conditions.

\section{Normal-Mode Lagrangian and Hamiltonian
	of the Coupled Magnon--Photon System}
\label{sec:finite_dim_NM_L_and_H}

\subsection{Normal-mode Lagrangian}
\label{sec:Polaritons_Lagrangian_NM}

In this section, we assume that the fields
$\Cvec_1$ and $\Mvec_1$ are described by
$P$ electromagnetic modes and $M$ magnonic modes, respectively. Accordingly,
unless otherwise specified, all summations are taken over
$p=1,\ldots,P$ and $m=1,\ldots,M$, where $p$ and $m$ denote the electromagnetic
and magnonic mode indices, respectively. The total number of modes is denoted by
\begin{align}
	N=M+P \, .
\end{align}

There are two reasons for introducing this finite-dimensional description.
First, normal-mode expansions are most useful when they lead to reduced-order
models involving only a limited number of electromagnetic and magnonic modes,
as exemplified by the few-degree-of-freedom descriptions underlying the
Jaynes--Cummings and Dicke models \cite{JaynesCummings1963,Dicke1954}.
Second, finite truncation allows the problem to be formulated in terms of
finite-dimensional matrices, considerably simplifying the notation. The
infinite-dimensional formulation is recovered by letting $P$ and $M$ tend to
infinity.

Thus, the dynamical state of the system is represented by the
finite-dimensional vector
\begin{align}
	\underline{z}
	=
	\left(
	\underline{b},\underline{b}^*;
	\underline{a},\underline{a}^*
	\right)
	\,\in\,\mathbb{C}^{2N} \, ,
	\label{eq:bosons_mag_phot}
\end{align}
where
\begin{align}
	&
	\underline{b}
	=
	(b_1,b_2,\ldots,b_M) \, ,
	\qquad
	\underline{b}^*
	=
	(b_1^*,b_2^*,\ldots,b_M^*) \, ,
	\\[3pt]
	&
	\underline{a}
	=
	(a_1,a_2,\ldots,a_P) \, ,
	\qquad
	\underline{a}^*
	=
	(a_1^*,a_2^*,\ldots,a_P^*) \, ,
\end{align}
denote the vectors of magnonic and electromagnetic normal-mode amplitudes, respectively. 
Throughout this section, all vectors are understood to
be column vectors.

It is worth emphasizing that, in the present formulation,
$\underline{z}^*$ is not regarded as an independent variable, but is instead
obtained from $\underline{z}$ through the linear relation
\begin{align}
	\underline{z}^*
	=
	C\cdot\underline{z} \, ,
	\label{eq:reL_z_z_conj}
\end{align}
where
\begin{align}
	C
	=
	\begin{pmatrix}
		0 & \mathbbm{I}_M & 0 & 0 \\[3pt]
		\mathbbm{I}_M & 0 & 0 & 0 \\[3pt]
		0 & 0 & 0 & \mathbbm{I}_P \\[3pt]
		0 & 0 & \mathbbm{I}_P & 0
	\end{pmatrix} \, .
	\label{eq:C_matrix}
\end{align}
Here, $\mathbbm{I}_M$ and $\mathbbm{I}_P$ denote the identity matrices of
dimensions $M\times M$ and $P\times P$, respectively. The matrix $C$
is a symmetric permutation matrix that acts as the complex-conjugation operator
only on vectors having the block structure given by
Eq.~\eqref{eq:bosons_mag_phot}.

Next, we derive the Lagrangian in terms of the normal-mode amplitudes.
The Lagrangians describing the free magnonic and free cavity-field dynamics
have already been obtained in
Eqs.~\eqref{eq:Lagrangian_bs} and \eqref{eq:Lagrange_ph_as}, respectively.
It therefore remains to express the interaction Lagrangian, namely the last
two terms in
Eq.~\eqref{eq:cavity_fild_lagrangian_coup_0_f}, in terms of the normal-mode
expansions~\eqref{eq:norma_modes_M1_expansion} and
\eqref{eq:expression_of_A_in_terms_of_electric_modes_a_a^*_f}.

The term describing the interaction between the magnetization and the cavity
is given by the right-hand side of
Eq.~\eqref{eq:identity_int_term_CM}. Substituting
Eqs.~\eqref{eq:norma_modes_M1_expansion} and
\eqref{eq:expression_of_A_in_terms_of_electric_modes_a_a^*_f} into this
expression, we obtain
\begin{align}
	\nonumber
	&L^\text{Z}_\text{int}
	= \\[2pt]
	&
	-\hbar \sum_{p,m}
	\frac{1}{\widetilde{\omega}_p}
	\left(
	(g^\text{Z}_{pm})^*\dot b_m
	+
	g_{pm}^\text{Z}\dot b_m^*
	\right)
	[a_p+a_p^*] ,
\end{align}
where
\begin{align}
	g_{pm}^\text{Z}
	=
	\sqrt{\frac{\gamma \mu_0 \Msat \widetilde{\omega}_p}{2}}\,
	\sqrt{\frac{|V_\text{M}|}{|V|}} O_{pm} \, ,
	\label{eq:gZ_definition}
\end{align}
and
\begin{align}
	O_{pm}
	=
	\frac{1}{|V_\text{M}|}
	\int_{V_\text{M}}
	\varphivec_m^*(\xvec)
	\cdot
	\hvec_p(\xvec)
	dV \, ,
	\label{eq:Z_overlap_integral}
\end{align}
is the corresponding overlap integral. The coupling coefficient
$g_{pm}^\text{Z}$ has the dimensions of an angular frequency, while
$\hbar g_{pm}^\text{Z}$ has the dimensions of energy and characterizes the
coupling between spin-wave mode $m$ and cavity electromagnetic mode $p$.

The term describing the modification of the magnetostatic field due to
cavity shielding, corresponding to the fifth term in
the Lagrangian~\eqref{eq:cavity_fild_lagrangian_coup_0_f}, becomes
\begin{align}
	\nonumber
	L^{\Delta}_\text{int}
	=
	-\frac{1}{2}\sum_{m,\ell}
	\Big[
	\hbar
	\left(
	g^{\Delta_1}_{\ell m} b_\ell b_m^*
	+
	(g^{\Delta_1}_{\ell m})^* b_\ell^* b_m
	\right)
	+ \\
	\quad
	\hbar
	\left(
	g^{\Delta_2}_{\ell m} b_\ell b_m
	+
	(g^{\Delta_2}_{\ell m})^* b_\ell^* b_m^*
	\right)
	\Big] \, ,
\end{align}
where
\begin{align}
	\label{eq:gDelta_sym_1}
	g^{\Delta_1}_{\ell m}
	=
	-\frac{\gamma\mu_0 \Msat}{|V_\text{M}|}
	\int_{V_\text{M}}
	\varphivec_\ell
	\cdot
	\Hvec_{1\Mvec}^\Delta[\varphivec_m^*]
	\, dV
	=
	(g^{\Delta_1}_{m\ell})^{*} \, , 
\end{align}
\begin{align}
	\label{eq:gDelta_sym_2}
\!\!\!\!\!\!	g^{\Delta_2}_{\ell m}
	=
	-\frac{\gamma \mu_0 \Msat}{|V_\text{M}|}
	\int_{V_\text{M}}
	\varphivec_\ell
	\cdot
	\Hvec_{1\Mvec}^\Delta[\varphivec_m]
	\, dV
	=
	g^{\Delta_2}_{m\ell} \, .
\end{align}
The symmetry relations follow from the symmetry of the operator
$\Hvec_{1\Mvec}^\Delta[\,\cdot\,]$
(see Appendix~\ref{sec:Helmholtz_Decomposition}).
The coefficients $g^{\Delta_1}_{\ell m}$ and $g^{\Delta_2}_{\ell m}$ have the
physical dimension of an angular frequency.
The corresponding contribution to the Hamiltonian is
$\mathcal{H}_\text{M}^{\Delta}=-L_\text{int}^{\Delta}$, which in matrix form reads
\begin{align}
	\mathcal{H}_\text{M}^{\Delta}
	=
	\frac{\hbar}{2}
	\begin{pmatrix}
		\underline{b}^{\,T} &
		(\underline{b}^{*})^{T}
	\end{pmatrix}
	\begin{pmatrix}
		G^{\Delta_2} & G^{\Delta_1} \\
		(G^{\Delta_1})^{*} & (G^{\Delta_2})^{*}
	\end{pmatrix}
	\begin{pmatrix}
		\underline{b} \\
		\underline{b}^{*}
	\end{pmatrix} \, ,
	\label{eq:shield_quad_form_modal}
\end{align}
where $G^{\Delta_1}$ and $G^{\Delta_2}$ are $M\times M $ matrices with entries:
\begin{align}
	(G^{\Delta_1})_{km}
	=
	g^{\Delta_1}_{km},
	\qquad
	(G^{\Delta_2})_{km}
	=
	g^{\Delta_2}_{km} \, .
	\label{eq:G^delta_matrices}
\end{align}
The quadratic form \eqref{eq:shield_quad_form_modal} is positive semidefinite, as shown in
Appendix~\ref{Sec:positive_H_delta_M}.

The complete normal-mode Lagrangian can therefore be written as
\begin{widetext}
	\begin{align}
		L\left(\underline{z},\underline{\dot{z}}\right)
		=
		\sum_{m}
		\frac{i\hbar}{2}
		\left(
		b_m^*\dot b_m
		-
		b_m\dot b_m^*
		\right)
		+
		\sum_{p}
		\frac{i\hbar}{2}
		\left(
		a_p^*\dot a_p
		-
		a_p\dot a_p^*
		\right)
		-
		\hbar
		\sum_{p,m}
		\frac{1}{\widetilde{\omega}_p}
		\left(
		(g^\text{Z}_{pm})^*\dot b_m
		+
		g_{pm}^\text{Z}\dot b_m^*
		\right)
		[a_p+a_p^*]
		-
		\mathcal{H}\left(\underline{z}\right) \, ,
		\label{eq:modal_lagrangian}
	\end{align}
\end{widetext}
where
\begin{align}
	\mathcal{H}(\underline{z})
	=
	\mathcal{H}_\text{M}\left(\underline{b},\underline{b}^*\right)
	+
	\mathcal{H}_\text{P}\left(\underline{a},\underline{a}^*\right)
	+
	\mathcal{H}_\text{M}^{\Delta}\left(\underline{b},\underline{b}^*\right)
	\label{eq:Hamiltonian_normal_mode_1}
\end{align}
is the Hamiltonian~\eqref{eq:Hamiltonian_lin_coupled_0_f_a} expressed in terms
of the normal-mode variables. In eq.~\eqref{eq:Hamiltonian_normal_mode_1},
$\mathcal{H}_\text{M}\left(\underline{b},\underline{b}^*\right)$ and
$\mathcal{H}_\text{P}\left(\underline{a},\underline{a}^*\right)$
denote the truncated forms of
eqs.~\eqref{eq:sw_energy_tris_0_3_f} and
\eqref{eq:Hamiltonian_modes_a_sym_f}, respectively, whereas
$\mathcal{H}_\text{M}^{\Delta}$ is the positive semidefinite
shielding-energy contribution.

Notice that the first three terms in
eq.~\eqref{eq:modal_lagrangian}, which are equal to
$L\left(\underline{z},\underline{\dot{z}}\right)
+\mathcal{H}\left(\underline{z}\right)$, provide the normal-mode representation
of the velocity-dependent quantity
$\Theta = \int_V \bm{\Pi}_{\Cvec_1}\cdot\partial_t\Cvec_1\,dV
+\int_{V_\text{M}}\bm{\Pi}_{\Mvec_1}\cdot\partial_t\Mvec_1\,dV$
appearing in the field-level Hamiltonian construction
\eqref{eq:Hamiltonian_C_pi_C}. Thus, the Zeeman interaction enters
the field formulation through $\Theta$. In the normal-mode description, this
velocity-dependent structure determines the Faddeev--Jackiw symplectic form,
as discussed in the next section.

\subsection{Faddeev--Jackiw derivation of the normal-mode Hamiltonian formulation}

To derive the equations of motion and the corresponding Hamiltonian
formulation, it is convenient to rewrite the Lagrangian
\eqref{eq:modal_lagrangian} in the first-order form
\begin{align}
	L\left(\underline{z},\underline{\dot{z}}\right)
	=
	[\underline{A}(\underline{z})]^T\cdot
	\dot{\underline{z}}
	-
	\mathcal{H}(\underline{z}) \, ,
	\label{eq:Faddeev_form}
\end{align}
where $\underline{A}(\underline{z})$ is a
$2N$-dimensional vector and $T$ denotes matrix transposition.
Equation~\eqref{eq:Faddeev_form} is the standard first-order Lagrangian
employed in the Faddeev--Jackiw formulation of Hamiltonian systems based on
the Poincar\'e--Cartan integral invariant~\cite{Faddeev1988}.

The corresponding Euler--Lagrange equations can be written in the compact
Hamiltonian form
\begin{align}
	F\cdot\dot{\underline z}
	=
	\nabla_{\underline{z}}\mathcal H,
	\label{eq:general_FJ_dynamics}
\end{align}
where the dot denotes a time derivative and the Faddeev--Jackiw matrix $F$ is
defined by
\begin{align}
	F_{rs}
	=
	\frac{\partial A_s}{\partial z_r}
	-
	\frac{\partial A_r}{\partial z_s} \, ,
\end{align}
with $r,s=1,\ldots,2N$. By construction, $F$ is skew-symmetric,
$F^T=-F$. Provided that $F$ is nonsingular,
eq.~\eqref{eq:general_FJ_dynamics} can be written in the standard Hamiltonian
form
\begin{align}
	\dot{\underline z}
	=
	J\cdot\nabla_{\underline z}\mathcal H \, ,
	\label{eq:Hamiltonian_form}
\end{align}
where
$J=F^{-1}$ is the Poisson matrix defining the generally noncanonical
Poisson-bracket structure \cite{Faddeev1988}.

In the present case, the vector
$\underline A(\underline z)$ depends linearly on the phase-space coordinates
and can therefore be written as
$\underline A(\underline z)=K\cdot\underline z$,
so that
$F=K^T-K$.
Hence, $F$ is a constant complex skew-symmetric matrix with the explicit block
form
\begin{align}
	F
	=
	\hbar
	\begin{pmatrix}
		0
		&
		-i\mathbbm{I}_M
		&
		\Gamma^{*}
		&
		\Gamma^{*}
		\\[8pt]
		i\mathbbm{I}_M
		&
		0
		&
		\Gamma
		&
		\Gamma
		\\[8pt]
		-\Gamma^\dagger
		&
		-\Gamma^T
		&
		0
		&
		-i\mathbbm{I}_P
		\\[8pt]
		-\Gamma^\dagger
		&
		-\Gamma^T
		&
		i\mathbbm{I}_P
		&
		0
	\end{pmatrix},
	\label{eq:F_FJ_matrix}
\end{align}
where $\dagger$ denotes the Hermitian transpose and $\Gamma$ is the
$M\times P$ matrix with elements
\begin{align}
	\Gamma_{mp}
	=
	{g^{\mathrm Z}_{pm}}/{\widetilde{\omega}_p} \, .
	\label{eq:gamma_zeeman_FJ}
\end{align}
The matrices $K$ and $F$ follow directly from
eq.~\eqref{eq:modal_lagrangian}, and $F$ is nonsingular for any coupling
matrix $\Gamma$, as shown in Appendix~\ref{sec:F_nonsingular}.

The Hamiltonian is a quadratic function of the phase-space coordinates and
can be written as
\begin{align}
	\mathcal H(\underline z)
	=
	\frac12\,
	\underline z^T\cdot
	D\cdot
	\underline z,
	\label{eq:Hamiltonian_quadratic_form}
\end{align}
where $D$ has the block structure
\begin{align}
	D
	=
	\hbar
	\begin{pmatrix}
		G^{\Delta_2}
		&
		\Omega_\mathrm{M}
		+
		G^{\Delta_1}
		&
		0
		&
		0
		\\[4pt]
		\Omega_\mathrm{M}
		+
		(G^{\Delta_1})^{*}
		&
		(G^{\Delta_2})^{*}
		&
		0
		&
		0
		\\[4pt]
		0
		&
		0
		&
		0
		&
		\Omega_\mathrm{P}
		\\[4pt]
		0
		&
		0
		&
		\Omega_\mathrm{P}
		&
		0
	\end{pmatrix},
	\label{eq:D_matrix_block_form}
\end{align}
with
\begin{align}
	\Omega_\mathrm{M}
	&=
	\mathrm{diag}
	(\omega_1,\ldots,\omega_M),
	\\
	\Omega_\mathrm{P}
	&=
	\mathrm{diag}
	(\widetilde{\omega}_1,\ldots,\widetilde{\omega}_P),
\end{align}
and where $G^{\Delta_1}$ and $G^{\Delta_2}$ are defined in
eq.~\eqref{eq:G^delta_matrices}.

Substituting eq.~\eqref{eq:Hamiltonian_quadratic_form} into
eq.~\eqref{eq:general_FJ_dynamics}, we obtain the linear equations of motion
\begin{align}
	F\cdot\dot{\underline{z}}
	=
	D\cdot\underline{z} \, .
	\label{eq:linear_hamiltonian_system}
\end{align}
It is important to note that the Zeeman interaction enters the equations of
motion through the structure of the Faddeev--Jackiw matrix $F$, via
eqs.~\eqref{eq:F_FJ_matrix} and \eqref{eq:gamma_zeeman_FJ}, whereas the
cavity-induced magnetostatic shielding enters explicitly through the
Hamiltonian matrix $D$ in eq.~\eqref{eq:D_matrix_block_form}.

In standard applications of the Faddeev--Jackiw formalism, the phase-space
variables are real, and consequently the matrices $F$ and $D$ are real.
Here, instead, $\underline{z}$ has complex components, and $F$ and $D$ are
in general complex. Nevertheless, the block structure of $\underline{z}$
(see eq.~\eqref{eq:bosons_mag_phot}) implies the symmetry relations
\begin{align}
	\label{eq:conj_sym_property}
	C\cdot F^{*}\cdot C
	=
	F \, ,
	\qquad
	C\cdot D^{*}\cdot C
	=
	D \, ,
\end{align}
where $C$ is defined in eq.~\eqref{eq:C_matrix}. Furthermore, $D$ is
symmetric, $D^T=D$, as follows from the Hermitian character of
$G^{\Delta_1}$ and the symmetry of $G^{\Delta_2}$ (see
eqs.~\eqref{eq:gDelta_sym_1} and \eqref{eq:gDelta_sym_2}), whereas $F$ is
skew-symmetric, $F^T=-F$. It then follows from
eq.~\eqref{eq:conj_sym_property} that
\begin{align}
	\label{eq:Hermitianit_CD_CF}
	(C\cdot D)^\dagger
	=
	C\cdot D \, ,
	\qquad
	(C\cdot F)^\dagger
	=
	-C\cdot F \, .
\end{align}
Thus, $C\cdot D$ is Hermitian, whereas $C\cdot F$ is skew-Hermitian.
For state vectors satisfying eq.~\eqref{eq:reL_z_z_conj}, the Hamiltonian
can therefore be written as
$\mathcal H(\underline z)=\frac12\,\underline z^{\dagger}\cdot
(C\cdot D)\cdot\underline z$,
which explicitly shows that it is real. Moreover, $C\cdot D$ is positive
definite. Indeed, the free magnetic and electromagnetic contributions are
positive definite, since all entries of $\Omega_\mathrm{M}$ and
$\Omega_\mathrm{P}$ are strictly positive, while the shielding contribution
\eqref{eq:shield_quad_form_modal} is positive semidefinite
(Appendix~\ref{Sec:positive_H_delta_M}).

\subsection{Diagonalization, Poisson brackets, and quantization}

We now determine the normal modes of the equations of motion
eq.~\eqref{eq:linear_hamiltonian_system}. To this end, we seek solutions of the form
\begin{align}
	\underline{z}(t)
	=
	\bar c_n\,
	\underline v_n\,
	e^{-i\Omega_n t} \, ,
	\label{eq:mode_evolution}
\end{align}
where $\Omega_n$ is the angular frequency of the $n$-th polariton normal mode,
$\underline v_n\in\mathbb{C}^{2N}$ is the corresponding eigenvector, and
$\bar c_n$ is a complex constant determined by the initial conditions.

Substituting
eq.~\eqref{eq:mode_evolution} into
eq.~\eqref{eq:linear_hamiltonian_system} yields the generalized eigenvalue problem
\cite{HornJohnson2013}
\begin{align}
	-i\Omega_n\,F\cdot\underline v_n
	=
	D\cdot\underline v_n \, ,
	\label{eq:polariton_eigenmodes}
\end{align}
which involves the matrix pair $F$ and $D$. Left-multiplying both sides of
eq.~\eqref{eq:polariton_eigenmodes} by $C$ gives the equivalent generalized
eigenvalue problem
$\Omega_n(-iC\cdot F)\cdot\underline v_n
=
C\cdot D\cdot\underline v_n$,
where both $-iC\cdot F$ and $C\cdot D$ are Hermitian matrices
(see eq.~\eqref{eq:Hermitianit_CD_CF}), and $C\cdot D$ is positive definite.
Therefore, by a standard argument
\cite[Sec.~7.6]{HornJohnson2013}, \cite{Colpa1978},
the eigenfrequencies $\Omega_n$ are real.

An important property of the spectrum is that the eigenpairs of
eq.~\eqref{eq:polariton_eigenmodes} always occur in pairs. Indeed, if
$(\Omega_n,\underline v_n)$ is an eigenpair, then
$\left(-\Omega_n,C\cdot\underline v_n^*\right)$
is also an eigenpair. This follows directly from the symmetry relations
eq.~\eqref{eq:conj_sym_property}, which imply that whenever
$\underline z(t)$ is a solution of
eq.~\eqref{eq:linear_hamiltonian_system},
$C\cdot\underline z^*(t)$ is also a solution.
Alternatively, the same result follows by taking the complex conjugate of
eq.~\eqref{eq:polariton_eigenmodes} and using
eq.~\eqref{eq:conj_sym_property}.

To simplify the notation, we introduce the convention
\begin{align}
	\underline v_{-\!n}
	=
	C\cdot\underline v_n^* \, ,
	\qquad
	\Omega_{-\!n}
	=
	-\Omega_n \, .
\end{align}

By appropriate manipulations of
eq.~\eqref{eq:polariton_eigenmodes}, we obtain
\begin{align}
	i(\Omega_k+\Omega_h)\,
	\underline v_h^T\cdot F\cdot\underline v_k
	=
	0 \, ,
\end{align}
with $k,h\in\{-N,\ldots,-1,1,\ldots,N\}$. Hence,
\begin{align}
	\label{eq:ortogonality}
	\underline v_h^T\cdot F\cdot\underline v_k
	=
	\underline v_h^T\cdot D\cdot\underline v_k
	=
	0,
	\qquad
	k\neq -h \, .
\end{align}

For $k=-h$, we choose the normalization such that
\begin{align}
	\label{eq:normalization_F}
	\underline v_h^T\cdot(-iF)\cdot\underline v_k
	&=
	\mathrm{sgn}(k)\,
	\hbar\,
	\delta_{k,-h} \, , \\
	\underline v_h^T\cdot D\cdot\underline v_k
	&=
	\hbar\Omega_{|k|}
	\delta_{k,-h} \, ,
	\label{eq:normalization_D}
\end{align}
with $k,h\in\{-N,\ldots,-1,1,\ldots,N\}$, where $\mathrm{sgn}(k)$ denotes the sign of $k$.

These orthogonality relations imply that the eigenvectors
$\underline v_k$ form a basis. Therefore,
\begin{align}
	\underline z(t)
	=
	\sum_k
	c_k(t)\,
	\underline v_k \, ,
	\label{eq:expansion_in_eigenvectors}
\end{align}
where $k\in\{-N,\ldots,-1,1,\ldots,N\}$ and
\begin{align}
	c_k(t)
	=
	\mathrm{sgn}(k)\hbar^{-1}\,
	\underline v_{-k}^T
	\cdot
	(-iF)
	\cdot
	\underline z(t) \, .
\end{align}
From this expression it follows that, for $n=1,\ldots,N$,
$c_{-n}=c_n^*$. Therefore,
eq.~\eqref{eq:expansion_in_eigenvectors} can be written as
\begin{align}
	\underline z(t)
	=
	\sum_{n=1}^N
	\left[
	c_n(t)\,
	\underline v_n
	+
	c_n^*(t)\,
	\underline v_{-n}
	\right] \, .
	\label{eq:expansion_polariton_cn}
\end{align}

Substituting this expansion into the Hamiltonian
eq.~\eqref{eq:Hamiltonian_quadratic_form} and using the normalization conditions
eqs.~\eqref{eq:ortogonality} and \eqref{eq:normalization_D}, we obtain
\begin{align}
	\label{eq:Hamiltonian_modes_polariton}
	\mathcal{H}
	=
	\sum_{n=1}^N
	\frac{\hbar\Omega_n}{2}
	\left(
	c_nc_n^*
	+
	c_n^*c_n
	\right)
	\, ,
\end{align}
which is the diagonal form of the polariton Hamiltonian.

Likewise, substituting
eq.~\eqref{eq:expansion_polariton_cn}
into
eq.~\eqref{eq:linear_hamiltonian_system}
and left-multiplying by $\underline v_{-n}^T$, we obtain
\begin{align}
	\label{eq:eqs_lin_dyn_classical_bosonic_c}
	\dot c_n
	=
	-i\Omega_n c_n \, ,
	\qquad
	\dot c_n^*
	=
	i\Omega_n c_n^* \, ,
\end{align}
confirming that $(c_n,c_n^*)$ are the normal-mode amplitudes of the coupled
system, namely, the magnetic-polariton amplitudes.

The symplectic structure of the problem could be inferred
indirectly by comparing
eqs.~\eqref{eq:Hamiltonian_modes_polariton}
and \eqref{eq:eqs_lin_dyn_classical_bosonic_c}. However, given
its fundamental role in the transition to the quantum description,
we derive it here directly.

First, we observe that the expansion
eq.~\eqref{eq:expansion_polariton_cn}
can be regarded as the linear change of variables
\begin{align}
	\underline{z}
	=
	V\cdot\underline{\psi} \, ,
	\label{eq:linear_changez_psi}
\end{align}
where
\begin{align}
	\label{eq:psi_vector}
	\underline{\psi}
	=
	(c_1,\ldots,c_N,c_1^*,\ldots,c_N^*) \, ,
\end{align}
and
\begin{align}
	V
	&=
	\left(
	\begin{array}{cccccc}
		\big| &        & \big| & \big| &        & \big| \\
		\underline v_1 & \cdots & \underline v_N
		& \underline v_{-\!1} & \cdots & \underline v_{-\!N} \\
		\big| &        & \big| & \big| &        & \big|
	\end{array}
	\right)
\end{align}
is the matrix whose columns are the generalized eigenvectors.
The transformation~\eqref{eq:linear_changez_psi} can be interpreted as a
generalized Bogoliubov transformation
\cite{Colpa1978,vanHemmen1980}.

From the Faddeev--Jackiw formulation
eq.~\eqref{eq:Faddeev_form}, the Poisson brackets of the dynamical
variables $z_r$ are given by
\begin{align}
	\{ z_r,z_s\}
	=
	F^{-1}_{rs} =J_{rs} \, .
	\label{eq:poisson_F_brackets}
\end{align}
Using the transformation
eq.~\eqref{eq:linear_changez_psi}
in
eq.~\eqref{eq:poisson_F_brackets}, we obtain
\begin{align}
	\label{eq:poisson_brackets_psi}
	\{ \psi_u,\psi_v\}
	=
	\left[
	V^{-1}\cdot F^{-1}\cdot V^{-T}
	\right]_{uv}
	\, ,
\end{align}
where $V^{-T}$ denotes the inverse transpose of $V$.

To determine the matrix on the right-hand side of
eq.~\eqref{eq:poisson_brackets_psi}, we write the eigenvalue problem
eq.~\eqref{eq:polariton_eigenmodes}
in the compact matrix form
\begin{align}
	-iF\cdot V\cdot\Omega
	=
	D\cdot V \, ,
	\label{eq:polariton_eigenproblem_matrix}
\end{align}
where
\begin{align}
	\Omega
	=
	\mathrm{diag}
	\left(
	\Omega_1,\ldots,\Omega_N,
	\Omega_{-1},\ldots,\Omega_{-N}
	\right) \, .
\end{align}
Using the normalization conditions
eqs.~\eqref{eq:normalization_F}
and \eqref{eq:normalization_D}
together with
eq.~\eqref{eq:polariton_eigenproblem_matrix},
we obtain after some algebra
\begin{align}
	V^{-1}\cdot F^{-1}\cdot V^{-T}
	=
	\frac{1}{i\hbar}
	\begin{pmatrix}
		0
		&
		+\mathbbm{I}_N
		\\
		-\mathbbm{I}_N
		&
		0
	\end{pmatrix}
	\, .
\end{align}
Substituting this result into
eq.~\eqref{eq:poisson_brackets_psi}
and using
eq.~\eqref{eq:psi_vector},
we finally obtain
\begin{align} 
	\label{eq:bosonic_PB_cs}
	\{c_n,c^*_\ell\}
	=
	\frac{\delta_{n\ell}}{i\hbar}
	\, ,
	\qquad
	\{c_n,c_\ell\}
	=
	\{c_n^*,c_\ell^*\}
	=
	0 \, ,
\end{align}
where $n,\ell=1,\ldots,N$.

The canonical quantization \cite{CohenTannoudji1989} of the polariton dynamics is then obtained by
promoting the classical amplitudes $(c_n,c_n^*)$ to bosonic operators
$(\hat c_n,\hat c_n^\dagger)$ satisfying the commutation relations
\begin{align}
	[\hat c_n,\hat c_\ell^\dagger]
	=
	\delta_{n\ell}\, ,
	\qquad
	[\hat c_n,\hat c_\ell]
	=
	[\hat c_n^\dagger,\hat c_\ell^\dagger]
	=
	0 \, ,
\end{align}
which follow from the correspondence principle
\begin{align}
	[\hat A,\hat B] \qquad
	\longleftrightarrow\qquad 
	i\hbar\,\{A,B\} \, .
\end{align}
Accordingly, the Hamiltonian operator becomes
\begin{align}
	\nonumber
	\hat{\mathcal H}
	=&
	\sum_{n=1}^{N}
	\frac{\hbar\Omega_n}{2}
	\left(
	\hat c_n\hat c_n^\dagger
	+
	\hat c_n^\dagger\hat c_n
	\right)
	= \\
	&\qquad \qquad \sum_{n=1}^{N}
	\hbar\Omega_n
	\left(
	\hat c_n^\dagger\hat c_n
	+
	\frac12
	\right),
\end{align}
which describes a set of independent bosonic polariton modes.

\section{Application to a cavity-magnonics problem}
\label{sec:application}

\subsection{Geometry and physical setting}
\label{sec:coupled_geometry}

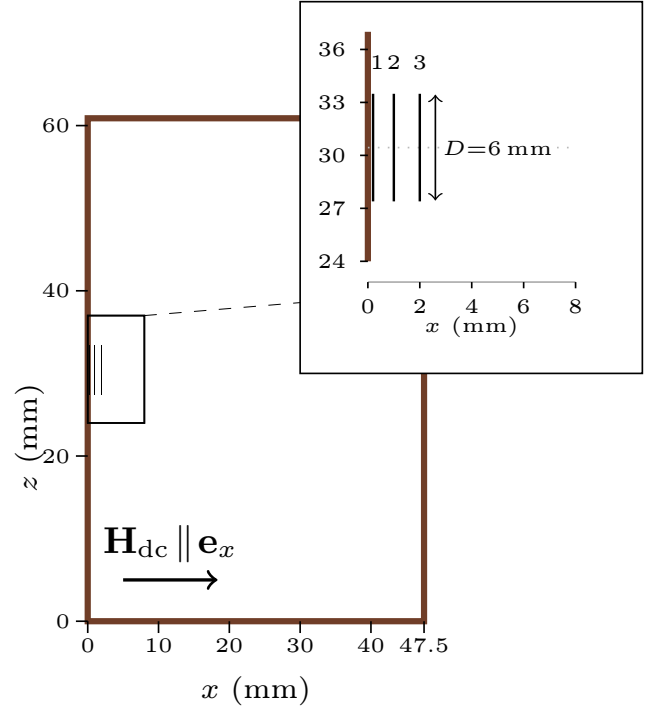
\begin{figure}[!t]
	\centering
	\resizebox{\columnwidth}{!}{%
		\begin{tikzpicture}[x=1cm,y=1cm]

			\draw[darkcopper,line width=1.5pt] (0,0) rectangle (2.85,4.262);
			
			\draw (0,0)--(0,-0.1);       \node[below=1pt,font=\tiny] at (0,0)     {0};
			\draw (0.6,0)--(0.6,-0.1);   \node[below=1pt,font=\tiny] at (0.6,0)   {10};
			\draw (1.2,0)--(1.2,-0.1);   \node[below=1pt,font=\tiny] at (1.2,0)   {20};
			\draw (1.8,0)--(1.8,-0.1);   \node[below=1pt,font=\tiny] at (1.8,0)   {30};
			\draw (2.4,0)--(2.4,-0.1);   \node[below=1pt,font=\tiny] at (2.4,0)   {40};
			\draw (2.85,0)--(2.85,-0.1); \node[below=1pt,font=\tiny] at (2.85,0)  {47.5};
			\node[below=10pt,font=\scriptsize] at (1.425,0) {$x$ (mm)};
			
			\draw (0,0)--(-0.1,0);       \node[left=1pt,font=\tiny] at (0,0)     {0};
			\draw (0,1.4)--(-0.1,1.4);   \node[left=1pt,font=\tiny] at (0,1.4)   {20};
			\draw (0,2.8)--(-0.1,2.8);   \node[left=1pt,font=\tiny] at (0,2.8)   {40};
			\draw (0,4.2)--(-0.1,4.2);   \node[left=1pt,font=\tiny] at (0,4.2)   {60};
			\node[left=14pt,rotate=90,font=\scriptsize] at (0,2.131) {$z$ (mm)};
			
			\draw[black,thin] (0,1.680) rectangle (0.480,2.590);
			
			\fill[black]
			(0.0111,1.9208) rectangle (0.0129,2.3408);
			\fill[black]
			(0.0591,1.9208) rectangle (0.0609,2.3408);
			\fill[black]
			(0.1191,1.9208) rectangle (0.1209,2.3408);
			
			\draw[->,line width=0.8pt] (0.3,0.35)--(1.1,0.35);
			\node[above=1pt] at (0.7,0.35)
			{$\Hvec_\mathrm{dc}\!\parallel\!\mathbf{e}_x$};
			
			\fill[white] (1.80,2.10) rectangle (4.70,5.25);
			\draw[black,thin] (0,1.680) rectangle (0.480,2.590);
			\draw[black,dashed,line width=0.15pt] (0.48,2.59)--(1.80,2.70);
			
			\begin{scope}[cm={0.44,0,0,0.44,(0.90cm,3.05cm)}]
				
				\draw[darkcopper,line width=1.5pt] (3.35,0)--(3.35,4.420);
				\draw[gray!65,thin] (3.35,-0.40)--(7.35,-0.40);
				
				\draw (3.35,-0.40)--(3.35,-0.52); \node[below=1pt,font=\tiny] at (3.35,-0.40) {0};
				\draw (4.35,-0.40)--(4.35,-0.52); \node[below=1pt,font=\tiny] at (4.35,-0.40) {2};
				\draw (5.35,-0.40)--(5.35,-0.52); \node[below=1pt,font=\tiny] at (5.35,-0.40) {4};
				\draw (6.35,-0.40)--(6.35,-0.52); \node[below=1pt,font=\tiny] at (6.35,-0.40) {6};
				\draw (7.35,-0.40)--(7.35,-0.52); \node[below=1pt,font=\tiny] at (7.35,-0.40) {8};
				\node[below=5pt,font=\tiny] at (5.35,-0.40) {$x$ (mm)};
				
				\draw (3.35,0.000)--  (3.20,0.000);  \node[left=2pt,font={\fontsize{4}{4}\selectfont}] at (3.35,0.000)  {24};
				\draw (3.35,1.020)--  (3.20,1.020);  \node[left=2pt,font={\fontsize{4}{4}\selectfont}] at (3.35,1.020)  {27};
				\draw (3.35,2.040)--  (3.20,2.040);  \node[left=2pt,font={\fontsize{4}{4}\selectfont}] at (3.35,2.040)  {30};
				\draw (3.35,3.060)--  (3.20,3.060);  \node[left=2pt,font={\fontsize{4}{4}\selectfont}] at (3.35,3.060)  {33};
				\draw (3.35,4.080)--  (3.20,4.080);  \node[left=2pt,font={\fontsize{4}{4}\selectfont}] at (3.35,4.080)  {36};
				
				\draw[gray!55,dotted] (3.35,2.190)--(7.35,2.190);
				
				\filldraw[fill=black]   (3.4425,1.170) rectangle (3.4575,3.210);
				\filldraw[fill=gray!60] (3.8425,1.170) rectangle (3.8575,3.210);
				\filldraw[fill=gray!25] (4.3425,1.170) rectangle (4.3575,3.210);
				
				\node[above=3pt,font=\tiny] at (3.530,3.210) {1};
				\node[above=3pt,font=\tiny] at (3.850,3.210) {2};
				\node[above=3pt,font=\tiny] at (4.350,3.210) {3};
				
				\draw[<->] (4.650,1.170)--(4.650,3.210);
				\node[right=1pt,font=\tiny,fill=white,inner sep=0.5pt]
				at (4.650,2.190) {$D{=}6\,\mathrm{mm}$};
				
			\end{scope}
			\draw[black,thin] (1.80,2.10) rectangle (4.70,5.25);
			
		\end{tikzpicture}
	}%
	
\caption{Cross-section ($xz$ plane, $y=b/2$) of the WR-187 cavity,
	with the three disk positions $x_c{=}0.2$, $1.0$, and
	$2.0\,\mathrm{mm}$, all centered at $(y_c,z_c)=(b/2,L/2)$. The inset
	magnifies the black-outlined region ($0\leq x\leq8\,\mathrm{mm}$,
	$24\leq z\leq37\,\mathrm{mm}$). The disk normal and the static bias
	field are along $x$.}
	\label{fig:coupled_geometry}
\end{figure}

We apply the theory developed above to a thin circular disk of yttrium iron
garnet (YIG), with diameter $D=6\,\mathrm{mm}$, thickness
$d=0.03\,\mathrm{mm}$, and saturation magnetization
$\mu_0 M_s=0.178\,\mathrm{T}$, magnetized along its symmetry axis and placed
inside a rectangular microwave cavity (Fig.~\ref{fig:coupled_geometry}), a
configuration extensively investigated in microwave physics
\cite{Dillon1960,YukawaAbe1974,Saha2001,Saha2002,Kamenetskii2004}
and, more recently, in cavity magnonics \cite{Flower2019}. The cavity
cross section, $a=47.5488\,\mathrm{mm}$ by $b=22.1488\,\mathrm{mm}$, is
that of a standard WR-187 waveguide \cite{Pozar2012}, and the length
$L=60.88\,\mathrm{mm}$ is chosen so that the fundamental TE$_{101}$ mode
resonates at $4\,\mathrm{GHz}$. The disk lies parallel to the side wall
$x=0$, near the magnetic-field maximum of this mode. To investigate the
influence of cavity shielding, its center is placed at $0.2$, $1$, or
$2\,\mathrm{mm}$ from the wall. Although the microwave relative
permittivity of YIG is about $15$--$16$ \cite{Bourhill2016}, dielectric
loading is negligible, because the sample-to-cavity volume ratio is of
order $10^{-5}$ and the disk lies close to a zero of the electric field of
the TE$_{101}$ mode.

Exchange is neglected, since $D,d\gg\lex$, and so is the weak
magnetocrystalline anisotropy of YIG, so that the uncoupled micromagnetic
normal modes are magnetostatic modes \cite{Walker1957}. We focus on the
fundamental branch of forward-volume magnetostatic waves in a
perpendicularly magnetized film \cite{StancilPrabhakar2009}, which has no
nodes across the thickness \cite{YukawaAbe1974,Charbois2002} and nearly
uniform dynamic magnetization through it (Sec.~I of the SM). This branch
contains the quasi-Kittel mode, the most spatially uniform and hence the
most strongly coupled to the cavity field. The modal description includes
the TE$_{101}$ mode, obtained analytically, and 15 modes of this branch,
computed with the micromagnetic eigensolver of Sec.~III of the SM and
chosen so as to include three modes with nonzero average magnetization:
the axisymmetric modes $m=1$, $6$, and $15$ of
Fig.~\ref{fig:first_fifteen_disk_modes_common_field}, labeled $(1,0)$,
$(3,0)$, and $(5,0)$ in Ref.~\cite{YukawaAbe1974}, where the first index,
denoted there by $n$, is the number of half-wavelengths of the radial
standing wave across a diameter (Sec.~I of the SM). From these modes the
coupling matrices are evaluated, and the hybridized modes are obtained
from the generalized eigenvalue problem~\eqref{eq:polariton_eigenmodes}
of the Faddeev--Jackiw formulation. As in a typical
ferromagnetic-resonance experiment, the static bias field is swept over a
range in which the relevant magnetic modes approach the $4\,\mathrm{GHz}$
cavity resonance while remaining well below the next one.

\subsection{Uncoupled normal modes}
\label{sec:uncoupled_modes}

\begin{figure}[!t]
	\centering
	\includegraphics[width=\columnwidth]{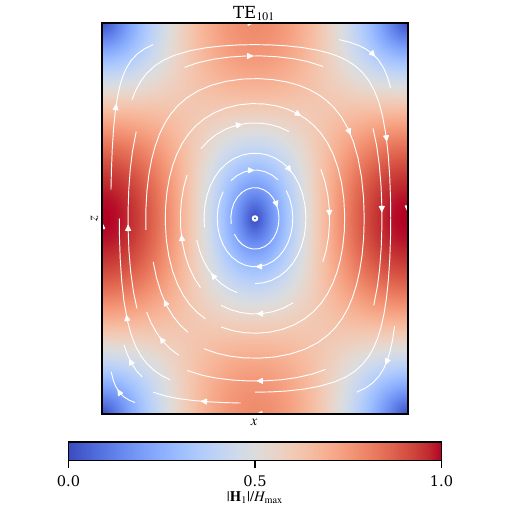}
	\caption{Magnetic field of the TE$_{101}$ mode ($4.00\,\mathrm{GHz}$)
		in the $xz$ plane of the empty WR-187 cavity. Color represents the
		normalized magnitude $|\mathbf H_1|/H_{\max}$, and the white
		streamlines show the direction of $\mathbf H_1$ at an instant of
		maximum magnetic amplitude. After half a period all directions
		reverse, while the field-line pattern and the magnitude are
		unchanged. The open circle marks the magnetic-field node at the
		cavity center.}
	\label{fig:wr187_te101}
\end{figure}

The normal modes of the empty rectangular cavity are labeled by the
collective index $p=(\sigma,p_x,p_y,p_z)$, where $\sigma=\mathrm{TE}$ or
$\mathrm{TM}$ specifies the mode family and $p_x$, $p_y$, $p_z$ are
nonnegative integers. With $k_x=p_x\pi/a$, $k_y=p_y\pi/b$,
$k_z=p_z\pi/L$, $k_\perp^2=k_x^2+k_y^2$, and
$k_{p_xp_yp_z}^2=k_\perp^2+k_z^2$, their frequencies are
$\widetilde\omega_{p_xp_yp_z}^\sigma=c k_{p_xp_yp_z}$, and the analytical
expressions of their magnetic fields are given in Sec.~II of the SM. The
fundamental mode TE$_{101}$ \cite{Pozar2012} has frequency
$\tilde{f}_{101}^\text{\,TE}=(c/2)\sqrt{a^{-2}+L^{-2}}=4\,\mathrm{GHz}$
for the chosen $L$, and the next mode, TE$_{102}$, lies at
$5.85\,\mathrm{GHz}$. The computations are performed in the band between
these two modes, where only the TE$_{101}$ mode needs to be retained. Its
magnetic field, shown in Fig.~\ref{fig:wr187_te101},
\begin{align}
	\label{eq:wr187_h101}
	\hvec_{101}^\text{TE}(\xvec)=\frac{2}{k_{101}}\big[&
	-k_z\sin(k_xx)\cos(k_zz)\,\evec_x \nonumber\\
	&+k_x\cos(k_xx)\sin(k_zz)\,\evec_z\big],
\end{align}
is independent of $y$ and lies in the $xz$ plane. It vanishes at the
cavity center ($x=a/2$, $z=L/2$), where the electric field is maximal,
and it is maximal at the centers of the two $yz$ walls ($x=0,a$,
$z=L/2$), where it is directed along $z$. It is normalized as in
eq.~\eqref{eq:orthogonality_relations},
$|V|^{-1}\int_V|\hvec_{101}^\text{TE}|^2\,dV=1$ with $|V|=abL$, and its
contribution to the electric vector potential and the associated
empty-cavity fields are
\begin{align}
	\label{eq:wr187_C101}
	\Cvec_1&=C_{101}(t)\,\hvec_{101}^\text{TE}(\xvec),
	\nonumber\\
	C_{101}(t)&=-\sqrt{\frac{\hbar}{2\mu_0|V|\widetilde\omega_{101}}}
	\left[a_{101}(t)+a_{101}^*(t)\right],
	\nonumber\\
	\Hvec_1&=-\dot C_{101}(t)\,\hvec_{101}^\text{TE}(\xvec),
	\nonumber\\
	\Evec_1&=-\frac{2k_{101}}{\varepsilon_0}\,C_{101}(t)
	\sin(k_xx)\sin(k_zz)\,\evec_y .
\end{align}
The three disk positions lie close to the magnetic-field maximum: at
$z=L/2$, $h_{101,z}(x)\propto\cos(\pi x/a)$, which at the three disk
centers equals $0.9999$, $0.9978$, and $0.9913$ times its maximum value
$h_{101,z}(0)$.

The magnetic modes are computed with a single cell across the disk
thickness, consistently with the restriction to the branch whose dynamic
magnetization is nearly uniform through it, on a grid of $140\times140$
cells in the $yz$ plane, with cell dimensions
$30\,\mu\mathrm{m}\times42.9\,\mu\mathrm{m}\times42.9\,\mu\mathrm{m}$
along $x$, $y$, and $z$. The circular cross section is represented by a
mask that selects the magnetic cells. The grid was refined until the
frequencies of all the modes considered changed by less than
$1\,\mathrm{MHz}$ between successive grids (Sec.~III of the SM).

\begin{figure*}[!t]
	\centering
	\includegraphics[width=0.98\textwidth]{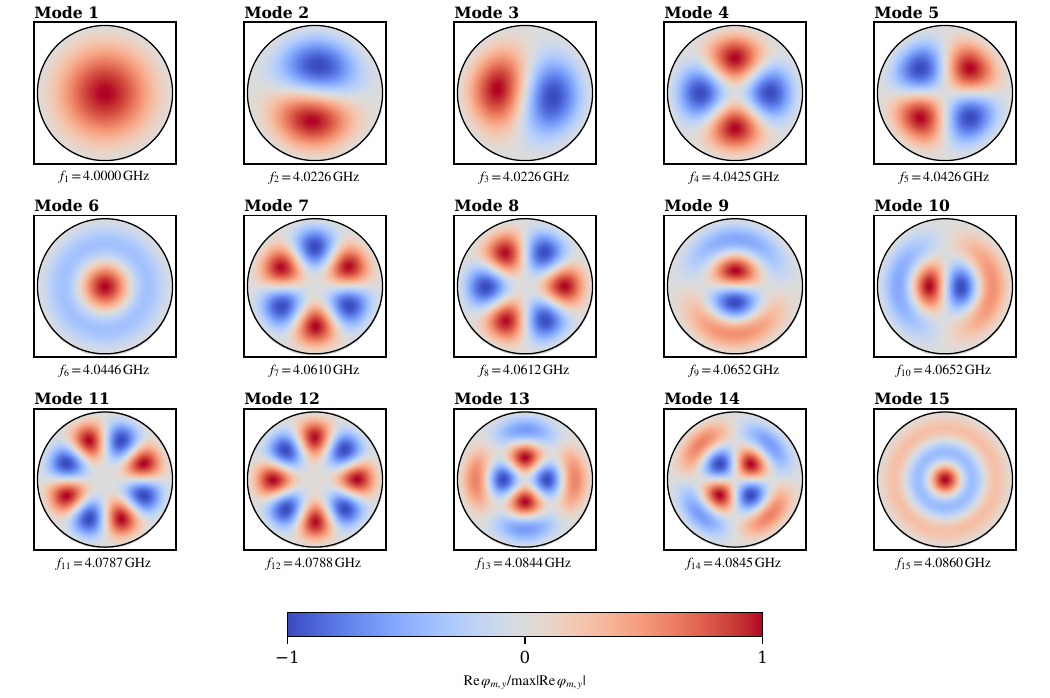}
\caption{First fifteen positive-frequency modes of the branch with nearly
	uniform dynamic magnetization across the thickness, for
	$\mu_0\Hvec_\text{dc}=(319.228\,\mathrm{mT})\,\evec_x$ ($140\times140\times1$
	mesh). Each panel shows the real part of the $y$ component of a mode,
	with the horizontal and vertical axes along $y$ and $z$, and the mode
	frequency below it. The slight splittings within nominally degenerate
	pairs are due to the discretization.}
	\label{fig:first_fifteen_disk_modes_common_field}
\end{figure*}

Since the bias field is along the disk axis and anisotropy is neglected,
the ground state is uniform, $\Mvec_0=\Msat\evec_x$, over the whole field
range considered. In the single-cell discretization it is an exact
equilibrium, because the in-plane demagnetizing field of a uniform normal
magnetization vanishes by symmetry, and the bias field enters the
linearized Landau--Lifshitz operator only as a uniform shift of $h_i$. The
modes are therefore computed once, at the center of the field sweep,
$\mu_0H_0=317.5\,\mathrm{mT}$, and the frequencies at the other fields are
$\omega_m(H_{\rm dc})=\omega_m(H_0)+\gamma\mu_0(H_{\rm dc}-H_0)(1+\varepsilon_m)$,
where
$\varepsilon_m=|V_\text{M}|^{-1}\int_{V_\text{M}}|\varphivec_m|^2\,dV-1$
accounts for the small ellipticity of the modes and, with the
normalization~\eqref{eq:Linearized_LL_modes_3a_f}, lies between
$1.8\times10^{-5}$ and $3.5\times10^{-4}$. The linearized operator is
represented in a local two-component basis transverse to $\Mvec_0$, and the
low-frequency modes are computed with a matrix-free implicitly restarted
Arnoldi method (ARPACK) with zero-shift inversion
\cite{Lehoucq1998,Saad2011}. The operator is applied through the
effective-field routines of standard finite-difference micromagnetics, with
the dipolar field evaluated by FFT convolution with the cell-averaged
demagnetizing tensor
\cite{DonahuePorter1999,Vansteenkiste2014MuMax3,dAquino_2009,dAquinoHertel2023}
(Sec.~III of the SM).

\begin{figure}[!t]
	\centering
	\includegraphics[width=\columnwidth]{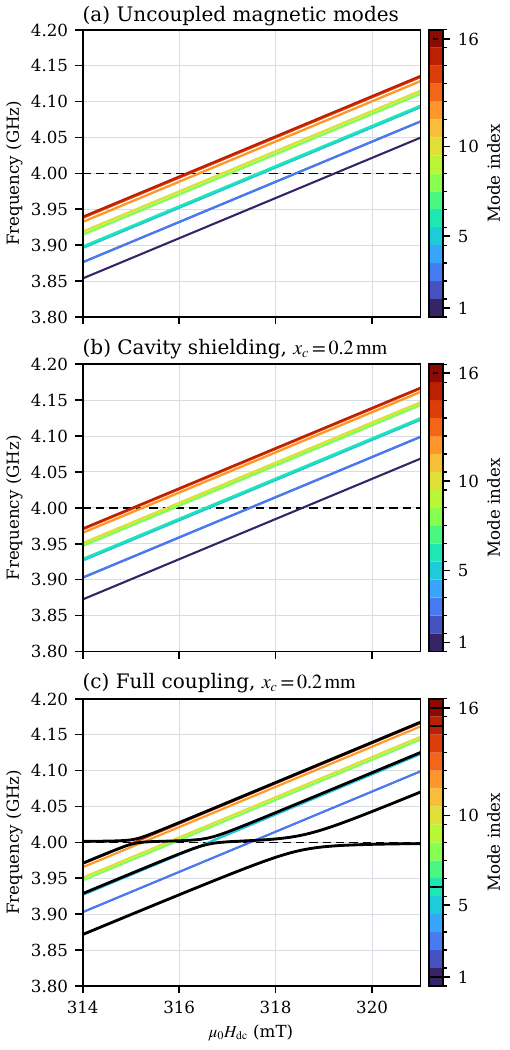}
\caption{Frequencies versus dc bias field ($140\times140\times1$ mesh).
	(a) Fifteen uncoupled magnetic modes. (b) Magnetic modes including the
	cavity-induced magnetostatic contribution at $x_c=0.2\,\mathrm{mm}$,
	without the direct Zeeman coupling. (c) Full magnon--photon spectrum at
	the same disk position, with the branches $n=1,6,15,16$ in black. The
	horizontal line in (a) and (b) marks the $4\,\mathrm{GHz}$ TE$_{101}$
	frequency.}
	\label{fig:first_fifteen_disk_modes_sweep}
\end{figure}

The quasi-Kittel mode ($m=1$) is resonant with the $4\,\mathrm{GHz}$
cavity frequency at $\mu_0H_{\rm dc}=319.228\,\mathrm{mT}$.
Figure~\ref{fig:first_fifteen_disk_modes_common_field} shows the fifteen
modes at this field, with their frequencies and the real part of
$\evec_y\cdot\varphivec_m(y,z)$. Over the sweep
$\mu_0 H_{\rm dc}=314$--$321\,\mathrm{mT}$ (50 equally spaced values),
the frequencies of the uncoupled modes,
Fig.~\ref{fig:first_fifteen_disk_modes_sweep}(a), follow from the linear
shift given above. Full calculations at $314$, $321$, and
$319.228\,\mathrm{mT}$, in which the equilibrium is obtained by relaxation
and the modes and coupling coefficients are recomputed at each field,
reproduce the magnetic and polariton frequencies, the coupling
coefficients, and the gaps discussed below to within $0.03\,\mathrm{MHz}$
(Sec.~III of the SM). All fifteen modes shift with the same slope,
$28.0\,\mathrm{MHz/mT}$, so that their spacings are preserved, and the
bright modes $m=1$, $6$, and $15$ cross the $4\,\mathrm{GHz}$ cavity
frequency at $\mu_0H_{\rm dc}=319.228$, $317.635$, and
$316.159\,\mathrm{mT}$, respectively.

\subsection{Coupled magnon--photon normal modes}
\label{sec:polariton_modes_results}

The coupled spectrum is obtained from the finite-dimensional eigenvalue
problem~\eqref{eq:polariton_eigenmodes} of
Sec.~\ref{sec:finite_dim_NM_L_and_H}, with the TE$_{101}$ mode as the only
photon mode ($P=1$) and the $M=15$ magnetic modes of
Fig.~\ref{fig:first_fifteen_disk_modes_common_field}, so that $N=16$.
Since the static fields are not affected by the cavity walls
(Sec.~\ref{sec:matter+field_lagrangian}), the ground state and the
uncoupled magnetic modes are the same for the three disk positions, and
they do not change with the bias field (Sec.~\ref{sec:uncoupled_modes}). The
coupling coefficients $g^{\mathrm Z}_{pm}$, $g^{\Delta_1}_{\ell m}$, and
$g^{\Delta_2}_{\ell m}$ are therefore computed once for each position. At
each of the 50 bias-field values they are combined with the magnetic
frequencies at that field to form the matrices $F$ and $D$
[eqs.~\eqref{eq:F_FJ_matrix} and \eqref{eq:D_matrix_block_form}], and the
generalized eigenvalue problem is solved. The polariton branches are
followed from one field value to the next by continuity and are labeled by
the uncoupled mode from which they originate, with $n=16$ denoting the
cavity mode.

The Zeeman coefficients $g^{\mathrm Z}_{pm}$ are given by the overlap
integral~\eqref{eq:gZ_definition} of each magnetic mode with the
TE$_{101}$ field, evaluated as a sum over the magnetic cells. Since the
cavity field is nearly uniform across the disk, only the axisymmetric
modes with nonzero average dynamic magnetization, $m=1$, $6$, and $15$,
couple appreciably. At $x_c=0.2\,\mathrm{mm}$,
$|g^{\mathrm Z}_{1m}|/2\pi\simeq 10.1$, $4.8$, and $3.2\,\mathrm{MHz}$,
respectively, while $|g^{\mathrm Z}_{1m}|/2\pi<0.03\,\mathrm{MHz}$ for all
other modes. The values at the other two positions differ only slightly,
because the TE$_{101}$ magnetic field varies weakly over this range. We
therefore call the magnetic branches $n=1,6,15$ and the cavity branch
$n=16$ \emph{bright}, and the remaining magnetic branches \emph{dark}.

The coefficients $g^{\Delta_1}_{\ell m}$ and $g^{\Delta_2}_{\ell m}$,
Eqs.~\eqref{eq:gDelta_sym_1}--\eqref{eq:gDelta_sym_2}, require the
cavity-induced field $\Hvec_{1\Mvec}^{\Delta}[\varphivec_m]$ of each mode,
fixed by the Neumann condition
$\nvec\cdot\Hvec_{1\Mvec}^{\Delta}=-\nvec\cdot\Hvec_\Mvec[\varphivec_m]$
on $\partial V$ (Eqs.~\eqref{eq:def_of_H1delta}--\eqref{eq:BC_for_psi}).
It is computed by a boundary-element method
\cite{Kress2014,SauterSchwab2011}, in which
$\Hvec_{1\Mvec}^{\Delta}$ is represented by a single-layer potential with
piecewise-constant density on rectangular wall panels, and the Neumann
condition, with the normal component of the free-space dipolar field of
the mode as datum, is imposed at the panel centers. The resulting dense
linear system, made solvable by the appropriate compatibility conditions,
is LU-factorized once and reused for all modes, and the overlap integrals
are computed by Gauss--Legendre quadrature over the magnetic cells
(Sec.~IV of the SM).

\begin{figure*}[!t]
	\centering
	\includegraphics[width=0.98\textwidth]{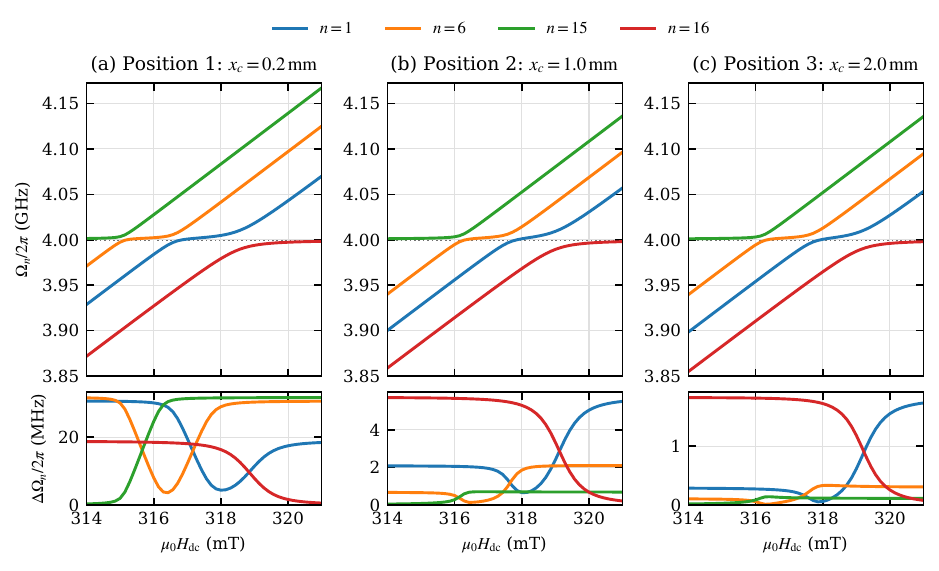}
	\caption{Cavity-shielding effect on the bright branches $n=1$, $6$,
		$15$, and $16$. Columns (a)--(c) correspond to the disk positions 1,
		2, and 3 of Fig.~\ref{fig:coupled_geometry} ($x_c=0.2$, $1.0$, and
		$2.0\,\mathrm{mm}$). The upper panels show the eigenfrequencies of
		the full calculation, including the direct Zeeman interaction and the
		cavity-induced magnetostatic contribution, with the same field and
		frequency ranges in the three columns. The dotted line marks the
		$4.00\,\mathrm{GHz}$ empty-cavity frequency. The lower panels show
		$\Delta\Omega_n/2\pi$ at every value of $H_{\rm dc}$, where
		$\Delta\Omega_n\equiv\Omega_n^{\mathrm{full}}-\Omega_n^{\mathrm{Z}}$
		and $\Omega_n^{\mathrm{Z}}$ is the eigenfrequency with the direct
		Zeeman interaction only, on vertical scales adjusted to each
		position.}
	\label{fig:shielding_effect_three_positions}
\end{figure*}

Figure~\ref{fig:first_fifteen_disk_modes_sweep}(b) shows the magnetic
spectrum at $x_c=0.2\,\mathrm{mm}$ with the shielding terms retained and
the direct Zeeman coupling omitted. Compared with
Fig.~\ref{fig:first_fifteen_disk_modes_sweep}(a), the shielding raises all
the magnetic branches, by a mode-dependent amount, and leaves the cavity
mode unchanged. The crossings with the $4\,\mathrm{GHz}$ cavity line
therefore move to lower fields, and they define the reference fields
$H_\times^{\Delta}$ at which the Zeeman-induced gaps are evaluated. In the
full spectrum, Fig.~\ref{fig:first_fifteen_disk_modes_sweep}(c), the
polariton branches follow the shielded magnetic modes and the empty-cavity
line away from the crossings. Avoided crossings, with gaps of the order of
the Zeeman coupling, appear only for the bright modes $m=1$, $6$, and $15$
(black lines), at progressively lower fields, while the dark branches cross
the cavity line with no visible gap.

Figure~\ref{fig:shielding_effect_three_positions} and
Table~\ref{tab:bright_mode_splittings} summarize the hybridized spectrum
for the three disk positions. The Zeeman-only gaps are essentially
independent of position (they vary by less than $1\%$), whereas the
shielding effect decreases rapidly away from the wall. At
$x_c=0.2\,\mathrm{mm}$ the shielding shifts the magnetic resonances by
about $20$--$30\,\mathrm{MHz}$, comparable to or larger than the gaps, and
displaces the crossing fields by up to $1.1\,\mathrm{mT}$. At
$x_c=2.0\,\mathrm{mm}$ the shifts fall below $2\,\mathrm{MHz}$. The effect
on the gaps is weaker, at most about $11\%$ at position 1 and $4\%$ at
positions 2 and 3. Cavity shielding therefore acts mainly as a
position-dependent shift of the resonance fields, with only a minor
correction to the coupling strength.

Dissipation is not included in the present model, but its relevance can be
assessed by comparing the coupling with the linewidths of the uncoupled
resonances. For $m=1$, the Zeeman-only gap of
Table~\ref{tab:bright_mode_splittings} gives
$g/2\pi=\Delta_{\rm Z}/2\simeq10\,\mathrm{MHz}$, or
$g/\widetilde\omega_{101}\simeq2.5\times10^{-3}$. Typical YIG damping,
$\alpha\simeq10^{-4}$--$5\times10^{-4}$
\cite{GurevichMelkov1996,Rezende2020,Rameshti2022}, gives at
$f=4\,\mathrm{GHz}$ a magnon linewidth
$\Gamma_{\rm m}/2\pi=2\alpha f\simeq0.8$--$4\,\mathrm{MHz}$. For the present
copper cavity the wall-loss limit is $Q\simeq1.1\times10^{4}$
\cite{Pozar2012}. Allowing for practical losses, we take
$Q=10^{3}$--$10^{4}$, i.e.,
$\Gamma_{\rm c}/2\pi=f/Q\simeq0.4$--$4\,\mathrm{MHz}$. The $m=1$ gap thus
exceeds either linewidth by at least a factor of five, and its
cooperativity, $C=4g^2/(\Gamma_{\rm m}\Gamma_{\rm c})$
\cite{Huebl2013,Zhang2014,HarderJAP2021}, ranges from about $25$ to more
than $10^3$, well within the strong-coupling regime. For $m=6$ and $15$,
$C$ is lower by factors of about $4$ and $10$, respectively. Although $C>1$
for both modes, this alone does not ensure a resolved splitting: the $m=6$
splitting should remain resolved for the assumed linewidths, whereas the
$m=15$ splitting is marginal at their upper end. Cavity shielding shifts the
resonance fields and modestly changes the gaps, without affecting the $m=1$
conclusion.

\section{Conclusions and perspectives}
\label{sec:conclusions}

We have developed a general Lagrangian and Hamiltonian normal-mode formulation
of cavity magnon--photon interactions, derived systematically from the coupled
Maxwell and Landau--Lifshitz equations linearized about an arbitrary
micromagnetic ground state. Rather than introducing the interaction through a
phenomenological coupled-mode Hamiltonian
\cite{SoykalFlatte2010,Huebl2013,Tabuchi2014,Zhang2014,Goryachev2014}, the
finite-dimensional description follows from the underlying field equations
after expansion in the normal modes of the uncoupled magnetic and
electromagnetic subsystems. A central point of the formulation is the
separation of the magnetization-generated magnetostatic field from the
radiative electromagnetic field. By introducing the electric vector potential
and an appropriate interaction Lagrangian, the magnetostatic contribution is
eliminated before the cavity-mode expansion is performed, thereby avoiding the
cancellation problem that arises in the conventional magnetic-vector-potential
formulation (Appendix~\ref{Sec:vector_pot_form_polaritons}).

\begin{table*}[!t]
\caption{Zeeman-induced gaps at the crossings of the bright magnetic modes
with the TE$_{101}$ mode, for the three disk positions. $\mu_0H_{\times}^{0}$
is the field at which the uncoupled magnetic mode crosses the cavity
frequency, and $\mu_0H_{\times}^{\Delta}$ the same crossing field with the
shielding terms retained and the direct Zeeman coupling omitted.
$\Delta_{\rm full}$ is the gap with both couplings at $H_{\times}^{\Delta}$,
and $\Delta_{\rm Z}$ the gap with the Zeeman coupling only at
$H_{\times}^{0}$, each being the separation of the two polariton modes with
the largest cavity participation. The last column gives
$100(\Delta_{\rm full}-\Delta_{\rm Z})/\Delta_{\rm Z}$. Crossing fields are
obtained by linear interpolation between the 50 computed field points.}
\label{tab:bright_mode_splittings}
\centering
\small
\setlength{\tabcolsep}{3.5pt}
\resizebox{0.88\textwidth}{!}{%
\begin{tabular}{ccccccc}
\hline\hline
Disk position & Mode $m$ & $\mu_0H_{\times}^{\Delta}$ (mT) & $\mu_0H_{\times}^{0}$ (mT) & $\Delta_{\rm full}$ (MHz) & $\Delta_{\rm Z}$ (MHz) & Change (\%) \\
\hline
1 ($x_c=0.2\,\mathrm{mm}$) & 1  & 318.556 & 319.228 & 20.783 & 20.119 & $+3.30$ \\
 & 6  & 316.540 & 317.635 & 8.717 & 9.558 & $-8.79$ \\
 & 15 & 315.028 & 316.159 & 5.840 & 6.545 & $-10.77$ \\
\hline
2 ($x_c=1.0\,\mathrm{mm}$) & 1  & 319.024 & 319.228 & 19.890 & 20.077 & $-0.93$ \\
 & 6  & 317.560 & 317.635 & 9.689 & 9.538 & $+1.58$ \\
 & 15 & 316.134 & 316.159 & 6.771 & 6.530 & $+3.68$ \\
\hline
3 ($x_c=2.0\,\mathrm{mm}$) & 1  & 319.163 & 319.228 & 19.794 & 19.947 & $-0.77$ \\
 & 6  & 317.624 & 317.635 & 9.673 & 9.476 & $+2.08$ \\
 & 15 & 316.155 & 316.159 & 6.602 & 6.486 & $+1.80$ \\
\hline\hline
\end{tabular}
}
\end{table*}

A novel aspect of the present formulation is the use of the Faddeev--Jackiw
approach \cite{Faddeev1988} to derive the Hamiltonian structure of the
coupled normal-mode system directly from its Lagrangian, which is of first
order in the generalized velocities. The resulting formulation is
noncanonical: the radiative magnon--photon interaction enters the
Poisson-bracket structure through geometry-dependent overlap coefficients,
whereas the cavity-induced modification of the magnetostatic field appears
explicitly in the Hamiltonian. The latter accounts for the shielding of the
magnetostatic field by the conducting cavity walls. Although this effect has
long been known in ferromagnetic-resonance studies
\cite{NashStandley1960,Srivastava1966}, it is often neglected in effective
cavity-magnonics models
\cite{LachanceQuirion2019,Flower2019,HarderJAP2021,Rameshti2022}. In the
YIG-disk example it mainly shifts the resonance fields, appreciably only
near the wall. The
same Hamiltonian structure leads naturally to the magnetic-polariton normal
modes through a generalized Bogoliubov transformation
\cite{Colpa1978,vanHemmen1980} and provides a basis for their subsequent
canonical quantization, along the lines of the quantum micromagnetic theory
of magnons in finite samples \cite{Mills2006,SerpicoPRB2025,Valet2026}.

The formulation can be extended to cavity geometries and electromagnetic
settings more general than those considered here. These include planar and
on-chip resonators \cite{Huebl2013,TabuchiScience2015,ClerkNatPhys2020},
dielectric loading, open boundaries, coupling ports, external driving, and
electromagnetic losses. In such cases, the electromagnetic modes generally
have to be determined numerically using standard methods of computational
electromagnetics, and for ports and open boundaries the closed-cavity
expansion must be supplemented by the corresponding port and radiation
terms \cite{Kurokawa1958}. These extensions mainly affect the
electromagnetic sector, while the micromagnetic normal-mode problem is
unchanged. Since the micromagnetic sector already includes exchange,
anisotropy, and arbitrary equilibrium states, the same procedure applies to
nonuniform and textured magnetic configurations
\cite{dAquino_2009,Valet2026}. The inclusion of dissipation would in
particular extend the formulation to driven--dissipative cavity-magnonic
systems, including dissipative coupling and non-Hermitian effects
\cite{Harder2018,HarderJAP2021,HurstFlebus2022,AshidaAdvPhys2020}.

A further extension concerns nonlinear magnetization dynamics and therefore
the magnetic sector itself. By carrying the expansion of the micromagnetic
energy and of the Landau--Lifshitz dynamics beyond quadratic order, the same
modal basis can be used to derive nonlinear coupling coefficients among
magnetic modes and between magnetic and electromagnetic modes
\cite{BertottiMayergoyzSerpico2009,PernaJMMM2022}. Such an extension would
provide a micromagnetic normal-mode description of nonlinear frequency
shifts, three- and four-magnon processes, Suhl instabilities
\cite{GurevichMelkov1996,BertottiMayergoyzSerpico2009}, and the nonlinear
dynamics of cavity magnon polaritons
\cite{Wang2016Kerr,Elyasi2020Nonlinear,Lee2023Nonlinear}.

\begin{acknowledgments}
	During the preparation of this work, the authors used generative
	artificial-intelligence tools. ChatGPT (OpenAI) and Claude (Anthropic) were
	used to assist with language editing and revision of the manuscript. Claude
	Code (Anthropic) and Codex (OpenAI) were used to assist in the development of
	the Fortran codes employed for the numerical simulations and of the Python
	scripts used for data analysis and figure generation, in running the
	numerical convergence and validation tests, and in drafting parts of
	the Supplemental Material. All material produced with the assistance
	of these tools was critically reviewed,
	verified, and edited by the authors, who take full responsibility for the
	content of the article.
\end{acknowledgments}
\appendix

\section{Helmholtz decomposition in $V$}
\label{sec:Helmholtz_Decomposition}

We consider the following Helmholtz decomposition result.
Any vector field $\vvec$ defined in the simply connected region $V$
can be uniquely decomposed into the sum of a divergence-free field
with vanishing normal component on $\partial V$ and a conservative field.
This result follows from the Tellegen-type identity
\begin{align}
	\label{eq:Tellegen_th}
	\int_V \uvec \cdot \nabla \psi \, dV = 0 \, ,
\end{align}
valid for any scalar field $\psi$, provided that
$\nabla \cdot \uvec = 0$ in $V$ and
$\nvec \cdot \uvec = 0$ on $\partial V$.
The uniqueness of the decomposition follows from
eq.~\eqref{eq:Tellegen_th}, since any harmonic field in $V$
with vanishing normal component on $\partial V$ must vanish identically.

In the case of the magnetization field $\Mvec$, the decomposition reads
\begin{align}
	\label{eq:HelmHoltz_decomp_V}
	\Mvec = \Bvec_\Mvec^V[\Mvec]/\mu_0 - \Hvec_\Mvec^V[\Mvec] \, ,
\end{align}
where $\Hvec_\Mvec^V[\Mvec]$ is the unique solution of the boundary-value
problem in
eqs.~\eqref{eq:Maxwell_for_B1M}--\eqref{eq:BC_on_BM^V}, while
$\Bvec_\Mvec^V[\Mvec]$
is a solenoidal field with vanishing normal component on $\partial V$.

An important consequence of the Helmholtz decomposition is the following
reciprocity theorem:
\begin{align}
	\label{eq:reciprocity_in_V}
	\int_V \Mvec_a \cdot \Hvec^V_\Mvec[\Mvec_b] \, dV
	=
	\int_V \Mvec_b \cdot \Hvec^V_\Mvec[\Mvec_a] \, dV \, ,
\end{align}
where $\Mvec_a$ and $\Mvec_b$ are two generic vector fields.
This result follows directly from
eqs.~\eqref{eq:Tellegen_th} and \eqref{eq:HelmHoltz_decomp_V}.

The magnetostatic field generated by a magnetized body in free space,
$\Hvec_\Mvec[\Mvec]$, defined in
eq.~\eqref{eq:Hm_operator_f}, also satisfies a reciprocity theorem
\cite{Brown1962}. It follows that
\begin{align}
	\label{eq:reciprocity_Delta_in_V}
	\int_V \Mvec_a \cdot \Hvec^\Delta_\Mvec[\Mvec_b] \, dV
	=
	\int_V \Mvec_b \cdot \Hvec^\Delta_\Mvec[\Mvec_a] \, dV \, ,
\end{align}
for generic vector fields $\Mvec_a$ and $\Mvec_b$.
We conclude that the operator $\Hvec^\Delta_\Mvec[\,\cdot\,]$ is
formally self-adjoint with respect to the standard $\mathbb{L}^2(V)$
inner product.

\section{Positivity of the operator $-{\Hvec}^\Delta_\Mvec[\Mvec_1]$}
\label{Sec:positive_H_delta_M}

We first observe that the boundary-value problem in
Eqs.~\eqref{eq:Maxwell_for_B1M}--\eqref{eq:BC_on_BM^V}
can be formulated in terms of a magnetic scalar potential
$\psi_\Mvec^V$, defined by
\begin{align}
	\label{eq:HM_as_function_of_psi}
	\Hvec_\Mvec^V = -\nabla \psi_\Mvec^V \, .
\end{align}
The potential $\psi_\Mvec^V$ satisfies Poisson's equation
\begin{align}
	\label{eq:poisson_psi_m_V}
	\nabla^2 \psi_\Mvec^V = \nabla \cdot \Mvec_1
\end{align}
in $V$, together with the boundary condition
\begin{align}
	\label{eq:Nuemann_homogenous_BC_for_psi}
	\nvec \cdot \nabla \psi_\Mvec^V = 0
\end{align}
on $\partial V$. Above and in the following, we assume that
$\Mvec_1$ is sufficiently smooth in order to avoid
unnecessary complications associated with the interface
conditions at the boundary of the magnetic body,
$\partial V_\text{M}$.

Next, we observe that the solution of the boundary-value problem
defined by
eqs.~\eqref{eq:poisson_psi_m_V} and
\eqref{eq:Nuemann_homogenous_BC_for_psi}
is the unconstrained minimum of the following Dirichlet functional:
\begin{align}
	V[\psi] = \int_V \frac{\mu_0}{2} |\nabla \psi|^2 \,dV +   \int_V \mu_0 \psi \nabla \cdot \Mvec_1 \,dV \, ,
\end{align}
namely,
\begin{align}
	\label{eq:arg_min_dirichlet}
	\psi_\Mvec^V = \operatorname*{arg\,min}_{\psi} V[\psi] \, .
\end{align}
We also note that
\begin{equation}
	V[ \psi_\Mvec^V ]= -W_\Mvec^V[\Mvec_1] = \frac{\mu_0}{2} \int_V \Mvec_1 \cdot \Hvec_\Mvec^V \, dV \, ,
\end{equation}
which is the negative of the magnetostatic energy
$W_\Mvec^V[\Mvec_1]$ associated with the field
$\Hvec_\Mvec^V$.

Now consider the case in which $\psi = \psi_\Mvec$, where
$\psi_\Mvec$ is the scalar potential satisfying
\begin{align}
	\Hvec_\Mvec = -\nabla \psi_\Mvec\, ,
\end{align}
where $\Hvec_\Mvec $ is defined in eq.~\eqref{eq:Hm_operator_f}.
The magnetostatic energy associated with this field is
\begin{align}
	\label{eq:energy_HM_in_V}
	W_\Mvec[\Mvec_1] = -\frac{\mu_0}{2} \int_{V} \Mvec_1 \cdot \Hvec_{\Mvec}[\Mvec_1]\, dV  ,
\end{align}
where the integral is taken over the entire cavity since
$\Mvec_1=0$ in the region $V\setminus V_\text{M}$.
The energy in Eq.~\eqref{eq:energy_HM_in_V}
can also be written as
\begin{align}
	\label{eq:energy_W_M_inft}
	W_\Mvec[\Mvec_1] = -\frac{\mu_0}{2} \int_V \psi_\Mvec\nabla \cdot \Mvec_1 \, dV \, .
\end{align}
Now, from Eqs.~\eqref{eq:arg_min_dirichlet} and
\eqref{eq:energy_W_M_inft}, it follows that
\begin{align}
	\nonumber
	& -W_\Mvec^V[\Mvec_1] \le \\ 
	\nonumber
	& \int_V \frac{\mu_0}{2} |\nabla \psi_\Mvec|^2 \,dV +   \int_V \mu_0 \psi_\Mvec \nabla \cdot \Mvec_1 \,dV \le \\
	& \int_{V_\infty} \frac{\mu_0}{2} |\nabla \psi_\Mvec|^2 \,dV - 2 W_\Mvec[ \Mvec_1]  \le -W_\Mvec[\Mvec_1] \, ,
\end{align}
where $V_\infty$ denotes the whole three-dimensional space.
Therefore,
\begin{align}
	W_\Mvec^V[\Mvec_1] \ge W_\Mvec[\Mvec_1] \, .
\end{align}
This implies that
\begin{align}
	\nonumber
	-\frac{\mu_0}{2} \int_{V} \Mvec_1 \cdot & \Hvec_{\Mvec}^\Delta[\Mvec_1]\, dV  = \\
	&	W_\Mvec^V[\Mvec_1] - W_\Mvec[\Mvec_1] \ge 0 \, ,
\end{align}
and, consequently, that the quadratic form associated with the operator
$-\Hvec_{\Mvec}^\Delta$ is positive semidefinite.

This positive magnetostatic energy correction is the energetic counterpart of
the wall effect mentioned at the end of Sec.~\ref{sec:Decomposition_of_B1}.

\section{Magnetic vector potential formulation}
\label{Sec:vector_pot_form_polaritons}

\subsection{Field equations}

Maxwell's equations, Eqs.~\eqref{eq:MaxwellrotE1_EB_coup_b}--\eqref{eq:MaxwelldivH1_EB_coup_b}, can be written as
\begin{align}
	\label{eq:MaxwellrotE1_EB_coup_b0}
	& \nabla \times \Evec_1 = -\partial_t \Bvec_1 \, , \\
	\label{eq:MaxwellrotB1_EB_coup_b0}
	& \nabla \times \Bvec_1 = \mu_0 \varepsilon_0 \partial_t \Evec_1 + \mu_0 \Jvec_{\Mvec_1} \, , \\
	\label{eq:MaxwelldivE1_EB_coup_b0}
	& \nabla \cdot \Evec_1 = 0 \, ,\\
	\label{eq:MaxwelldivH1_EB_coup_b0}
	& \nabla \cdot \Bvec_1 = 0 \, ,
\end{align}
where $\Jvec_{\Mvec_1}=\nabla \times \Mvec_1$. The LL equation takes the form
\begin{align}
	\label{eq:general_VLL_Gqs_lin_B}
	\partial_t \Mvec_1
	=
	\gamma \Mvec_0 \times
	\left(
	\delta_{\Mvec_1}\mathcal{H}_\text{M}
	-
	\Bvec_1
	+
	\Bvec_\Mvec[\Mvec_1]
	\right)
	\, ,
\end{align}
where $\Bvec_\Mvec[\Mvec_1] = \mu_0\left(\Hvec_\Mvec[\Mvec_1] + \Mvec_1\right)$.

Throughout this section, including in the equations above, we assume that
$\Mvec_1$ is a sufficiently smooth vector field to avoid
unnecessary technicalities associated with the electromagnetic interface
conditions.

We introduce the magnetic vector potential $\Avec_1$ through
\begin{align}
	\label{eq:Coulomb_Gauge_conds_1}
	\nabla \times \Avec_1
	&= \Bvec_1
	\, , \quad
	\nabla \cdot \Avec_1
	=
	0
	\qquad
	\text{in } V \, ,\\[2pt]
	\label{eq:bc_for_A1}
	&
	\nvec \times \Avec_1
	=
	0
	\qquad
	\text{on } \partial V \, .
\end{align}
Substituting these expressions into eq.~\eqref{eq:MaxwellrotE1_EB_coup_b0}, we obtain
\begin{align}
	\label{eq:E1_and_vector_pot_A1}
	\Evec_1 = -\partial_t \Avec_1 \, .
\end{align}
Equations \eqref{eq:Coulomb_Gauge_conds_1}, \eqref{eq:bc_for_A1},
and \eqref{eq:E1_and_vector_pot_A1} automatically satisfy
Eqs.~\eqref{eq:MaxwellrotE1_EB_coup_b0},
\eqref{eq:MaxwelldivE1_EB_coup_b0},
and \eqref{eq:MaxwelldivH1_EB_coup_b0},
as well as the boundary condition
in eq.~\eqref{eq:E_boundary_cond_E1}.

Accordingly, the vector potential $\Avec_1$ belongs to the linear space
\begin{align}
	\nonumber
	&	\mathbb{H}_{\Avec}^{\mathrm{em}}(V)
	=
	\big\{
	\wvec(\xvec)
	\,:\,
	\nabla\cdot\wvec=0
	\,\, \text{in }V \,, \\
	&   \quad
	\nvec\times\wvec=0
	\,\, \text{on }\partial V \,  , \,
	\{\wvec,\nabla\times\wvec\}\in\mathbb{L}^2(V)
	\big\}
	\, .
	\label{eq:vector_potential_space_A}
\end{align}
The Lagrangian associated with these evolution equations is
\begin{align}
	\nonumber
	&	{\mathcal L}
	=
	\int_V
	\left[
	\frac{\varepsilon_0}{2}
	(\partial_t\Avec_1)^2
	-
	\frac{(\nabla\times\Avec_1)^2}{2\mu_0}
	\right]
	dV
	+L_\text{M}[\Mvec_1,\partial_t \Mvec_1] \\
	&		-	\int_{V_\mathrm{M}}
	\frac12
	\Mvec_1\cdot
	\Bvec_\Mvec[\Mvec_1]
	\,dV
	+
	\int_V
	\Avec_1\cdot \Jvec_{\Mvec_1}
	\,dV
	\, .
	\label{eq:cavity_fild_lagrangian_coup_0_f_A}
\end{align}

The Lagrangian~\eqref{eq:cavity_fild_lagrangian_coup_0_f_A}
decomposes into a field part (the first integral), a magnetic part
$L_\text{M}[\Mvec_1, \partial_t\Mvec_1]$, and an interaction part (the last two terms).
The magnetic part coincides with that in
eq.~\eqref{eq:cavity_fild_lagrangian_coup_0_f}: in both formulations, the
unperturbed magnetic dynamics is that of the body without the cavity and is
governed by $\mathcal{H}_\text{M}$. The interaction, however, contains, in
addition to the bilinear term, the quadratic term
$-\frac12\int_{V_\mathrm{M}}\Mvec_1\cdot\Bvec_\Mvec[\Mvec_1]\,dV$. As will be
discussed below, its contribution to the frequencies of the hybridized modes is
not small.

The functional derivatives of the Lagrangian with
respect to $\partial_t\Avec_1$ and $\partial_t\Mvec_1$ are
\begin{align}
	\label{eq:conj_momentum_A1_Aform}
	& \bm{\Pi}_{\Avec_1}
	= \frac{\delta \mathcal{L}}{\delta (\partial_t \Avec_1)}
	= \varepsilon_0 \partial_t \Avec_1 \, , \\[3pt]
	& \bm{\Pi}_{\Mvec_1}
	= \frac{\delta \mathcal{L}}{\delta (\partial_t \Mvec_1)}
	= \frac{1}{2\gamma\Msat^2}
	\Mvec_0 \times \Mvec_1 \, ,
	\nonumber
\end{align}
and are used to construct the corresponding Hamiltonian:
\begin{align}
	\nonumber
	& \mathcal{H} =
	\int_V \!\left[
	\frac{(\bm{\Pi}_{\Avec_1})^2}{2\varepsilon_0}
	+ \frac{(\nabla \times \Avec_1)^2}{2\mu_0}
	\right] dV  \\
	\nonumber
	& \qquad  +\mathcal{H}_\text{M}[\Mvec_1]	+ \int_{V_\text{M}} \frac{1}{2} \Mvec_1 \cdot \Bvec_\Mvec[\Mvec_1] \, dV \\
	&
	\qquad\qquad\qquad	- \int_V (\nabla \times \Avec_1) \cdot \Mvec_1 \, dV		\, ,
	\label{eq:Hamiltonian_lin_coupled_Aform}
\end{align}
where the first integral is the electromagnetic field Hamiltonian,
$\mathcal{H}_\text{M}[\Mvec_1]$ is the magnetic Hamiltonian, and the last two
terms form the interaction Hamiltonian.

The dynamical variables $\Avec_1$ and $\bm{\Pi}_{\Avec_1}$ are constrained
to belong to the space $\mathbb{H}_{\Avec}^\text{em}(V)$,
while $\Mvec_1$ is constrained to belong to $\mathbb{H}^\text{sw}(V_\text{M})$.

The Euler--Lagrange equation for $\Avec_1$, obtained from the Lagrangian
\eqref{eq:cavity_fild_lagrangian_coup_0_f_A} under the constraint
$\Avec_1 \in \mathbb{H}_{\Avec}^{\mathrm{em}}(V)$, is
\begin{align}
	\label{eq:A1evolution_1_coup_A0}
	c^{-2}\partial_t^2\Avec_1
	-
	\nabla^2\Avec_1
	=
	\mu_0\Jvec_{\Mvec_1} \, ,
\end{align}
which coincides with
eq.~\eqref{eq:MaxwellrotB1_EB_coup_b0}
expressed in terms of $\Avec_1$.
The Euler--Lagrange equation for $\Mvec_1$, under the constraint
$\Mvec_1 \in \mathbb{H}^\text{sw}(V_\text{M})$, is
\begin{align}
	\nonumber
	\partial_t\Mvec_1
	&=
	\gamma\Mvec_0\times
	\\
	&
	\left(
	\delta_{\Mvec_1}\mathcal{H}_\text{M}
	-
	\nabla\times\Avec_1
	+
	\Bvec_\Mvec[\Mvec_1]
	\right)
	\, ,
	\label{eq:general_VLL_Gqs_lin_b0}
\end{align}
which is the constitutive relation
in eq.~\eqref{eq:general_VLL_Gqs_lin_B}
expressed in terms of $\Avec_1$.

The corresponding Hamiltonian equations follow by substituting the expression
for the momentum~\eqref{eq:conj_momentum_A1_Aform}
into eq.~\eqref{eq:A1evolution_1_coup_A0}; Eq.~\eqref{eq:general_VLL_Gqs_lin_b0}
is unchanged in the Hamiltonian formulation.

Note that $\Bvec_1=\mu_0\Hvec_\text{1R}+\Bvec_\Mvec^V[\Mvec_1]$ contains
the magnetization-generated field $\Bvec_\Mvec^V[\Mvec_1]$, which is
concentrated around the magnetic body and has a discontinuous tangential
component across $\partial V_\text{M}$. When we expand
$\Bvec_1=\nabla\times\Avec_1$ in terms of solenoidal cavity modes,
the low-order cavity modes retained in a reduced description vary on the
scale of the cavity, typically centimeters, whereas the magnetic body has
millimeter or submillimeter dimensions. Since the modal coefficients of
$\Bvec_\Mvec^V[\Mvec_1]$ are the projections of $\Mvec_1$ onto the cavity
modes,
$\Bvec_\Mvec^V[\Mvec_1]=(\mu_0/|V|)\sum_p\hvec_p\int_V\hvec_p\cdot\Mvec_1\,dV$,
an accurate representation requires a large number of modes, with wavelengths
extending down to the dimensions of the body.

On the other hand, eq.~\eqref{eq:general_VLL_Gqs_lin_b0} does not involve
$\nabla\times\Avec_1$ alone, but only the combination
\begin{align}
	\nabla\times\Avec_1-\Bvec_\Mvec[\Mvec_1]=
	\mu_0\left(\Hvec_\text{1R}+\Hvec^\Delta_{1\Mvec}[\Mvec_1]\right) ,
	\label{eq:A_LL_field_combination}
\end{align}
which is the field that drives the magnetization in
eq.~\eqref{eq:general_VLL_Gqs_lin_bis}. In the exact theory, the two
formulations are therefore equivalent. They differ only when the cavity field
is represented by a finite number $P$ of modes.
The difficulties arising from this truncation are discussed
in more detail in the next subsection.

To avoid the difficulty associated with the magnetic-vector-potential
formulation, one could introduce the decomposition
\begin{align}
	\Avec_1
	=
	\Avec_{\mathrm{1R}}
	+
	\Avec_\Mvec[\Mvec_1] \, ,
	\label{eq:decomp_Avec}
\end{align}
where
$\nabla\times\Avec_\Mvec[\Mvec_1]=\Bvec_\Mvec^V[\Mvec_1]$ (see eq.~\eqref{eq:HelmHoltz_decomp_V}),
$\nabla\cdot\Avec_\Mvec[\Mvec_1]=0$,
and $\Avec_\Mvec[\Mvec_1]$ satisfies the same boundary conditions as
$\Avec_1$.
With this decomposition, the Landau--Lifshitz equation depends only on the radiative
component $\Avec_{\mathrm{1R}}$ and the harmonic field
$\mu_0\Hvec^\Delta_{1\Mvec}[\Mvec_1]=\Bvec_\Mvec^V[\Mvec_1]-\Bvec_\Mvec[\Mvec_1]$,
thereby eliminating the cancellation problem.
However, substituting
eq.~\eqref{eq:decomp_Avec}
into the Lagrangian
eq.~\eqref{eq:cavity_fild_lagrangian_coup_0_f_A}
produces a considerably more complicated functional.
In particular, the electromagnetic kinetic energy generates additional terms
containing $\partial_t\Mvec_1$, resulting in a Lagrangian with a more intricate
symplectic structure and a correspondingly more involved Hamiltonian formulation.

\subsection{Normal-mode analysis with $M=1$ and arbitrary $P$ in the
	$\Avec$-formulation}

We consider the disk of Sec.~\ref{sec:application} at $x_c=2\,\mathrm{mm}$ from
the wall, a single micromagnetic mode, and an arbitrary number $P$ of cavity
modes.

The magnetic mode is the quasi-Kittel mode $\varphivec_1$, i.e., the
lowest-frequency micromagnetic normal mode of the disk. At
$\mu_0H_\mathrm{dc}=319.143\,\mathrm{mT}$ its angular frequency is
$\omega_1/2\pi=3.99761\,\mathrm{GHz}$. Denoting its complex amplitude by
$b(t)$, the magnetization is, by eq.~\eqref{eq:norma_modes_M1_expansion},
\begin{equation}
	\Mvec_1=\sqrt{\frac{\gamma\hbar\Msat}{|V_\text{M}|}}
	\left[b(t)\varphivec_1+b^*(t)\varphivec_1^*\right] .
	\label{eq:M1_single_mode}
\end{equation}

The cavity modes are labeled by $p=1,2,3,\dots$ in order of increasing
angular frequency $\widetilde\omega_p$; the mode $p=1$ is the TE$_{101}$
mode, with $\widetilde\omega_1/2\pi=4.000\,\mathrm{GHz}$. Since
$\Avec_1\in\mathbb{H}_{\Avec}^{\mathrm{em}}(V)$, it is expanded in the
electric-field modes
\begin{align}
	\evec_p=\frac{1}{k_p}\nabla\times\hvec_p \, ,
	\label{eq:electric_modes_from_h}
\end{align}
which, by
eqs.~\eqref{eq:electric_sol_modes_V_0_f}--\eqref{eq:orthogonality_relations},
are real, divergence-free, satisfy $\nvec\times\evec_p=0$ on $\partial V$ and
$\nabla\times\evec_p=k_p\hvec_p$, and are orthonormal,
$|V|^{-1}\int_V\evec_p\cdot\evec_{p'}\,dV=\delta_{pp'}$. Following the same
steps as in Sec.~\ref{sec:normal_modes_non_int}, the vector potential is
written in terms of the complex amplitudes $a_p(t)$ of the cavity modes as
\begin{align}
	\Avec_1(\xvec,t)
	=
	\sum_{p\le P}
	\sqrt{\frac{\hbar}{2\varepsilon_0|V|\,\widetilde\omega_p}}
	\left[a_p(t)+a_p^*(t)\right]\evec_p(\xvec) \, ,
	\label{eq:A1_modal_expansion}
\end{align}
so that the free cavity-field Hamiltonian, the first integral in
eq.~\eqref{eq:Hamiltonian_lin_coupled_Aform}, becomes
$\sum_p\hbar\widetilde\omega_p|a_p|^2$.
Since $\Bvec_1=\nabla\times\Avec_1$ and $\nabla\times\evec_p=k_p\hvec_p$, the
quantity $a_p+a_p^*$ is proportional to the amplitude of the total magnetic flux
density of mode $p$, including the contribution of the field generated by the
magnetization. The amplitudes $a_p$ are therefore not the same variables as
those of eq.~\eqref{eq:expression_of_A_in_terms_of_electric_modes_a_a^*_f},
where $a_p+a_p^*$ is proportional to the electric field and $\Hvec_\text{1R}$
contains only the radiative part.

Each cavity mode $p$ is coupled to the magnetic mode through the Zeeman
coupling coefficient $G_p=g^\text{Z}_{p1}$, defined in
eq.~\eqref{eq:gZ_definition}, which has the dimension of an angular
frequency. For instance, for the TE$_{101}$ mode,
$|G_1|/2\pi=10.04\,\mathrm{MHz}$.

Two further constants appear in the analysis below. The first is the
frequency shift produced by the quadratic term
$-\frac12\int_{V_\text{M}}\Mvec_1\cdot\Bvec_\Mvec[\Mvec_1]\,dV$ present both in
the Lagrangian~\eqref{eq:cavity_fild_lagrangian_coup_0_f_A} and in the
Hamiltonian~\eqref{eq:Hamiltonian_lin_coupled_Aform}. Substituting
eq.~\eqref{eq:M1_single_mode}, its normal part contributes
$\hbar\omega_Q|b|^2$ to the Hamiltonian, with
\begin{equation}
	\omega_Q=\frac{\gamma\mu_0\Msat}{|V_\text{M}|}\int_{V_\text{M}}
	\varphivec_1^*\cdot\frac{\Bvec_\Mvec[\varphivec_1]}{\mu_0}\,dV .
\end{equation}
For the computed quasi-Kittel mode, $\omega_Q/2\pi=4.9701\,\mathrm{GHz}$,
close to $\gamma\mu_0\Msat/2\pi=4.989\,\mathrm{GHz}$.
The second constant is the shift produced by the shielding term
$\frac{\mu_0}{2}\int_{V_\text{M}}\Mvec_1\cdot\Hvec^\Delta_{1\Mvec}[\Mvec_1]\,dV$
of the Lagrangian~\eqref{eq:cavity_fild_lagrangian_coup_0_f}, whose normal part
contributes $\hbar\omega_\Delta|b|^2$, with
\begin{equation}
	\omega_\Delta=-\frac{\gamma\mu_0\Msat}{|V_\text{M}|}\int_{V_\text{M}}
	\varphivec_1^*\cdot\Hvec^\Delta_{1\Mvec}[\varphivec_1]\,dV
	=g^{\Delta_1}_{11} \, ,
\end{equation}
where $g^{\Delta_1}_{11}$ is defined in eq.~\eqref{eq:gDelta_sym_1};
numerically, $\omega_\Delta/2\pi=1.831\,\mathrm{MHz}$.
The anomalous parts of the two terms, proportional to $b^2$ and $b^{*2}$, have
coefficients proportional to
$\int_{V_\text{M}}\varphivec_1\cdot\Bvec_\Mvec[\varphivec_1]\,dV$ and
$g^{\Delta_2}_{11}$ (see eq.~\eqref{eq:gDelta_sym_2}), respectively. The first
vanishes owing to the rotational symmetry of the disk and of the ground state,
together with the nondegeneracy of the quasi-Kittel mode (numerically, it is
below $10^{-8}\,\omega_Q$). The second is nonzero, because the cavity walls
break this symmetry, but it is small,
$|g^{\Delta_2}_{11}|/2\pi\simeq0.01\,\mathrm{MHz}$, and is neglected.

The constants $\omega_Q$ and $\omega_\Delta$ and the Zeeman couplings
$G_p$ are connected by the sum rule
\begin{align}
	\sum_{p=1}^{\infty}w_p=\omega_Q-\omega_\Delta \, ,
	\label{eq:sumrule}
\end{align}
where
\begin{align}
	w_p=\frac{2|G_p|^2}{\widetilde\omega_p}
	=\gamma\mu_0\Msat\,\frac{|V_\text{M}|}{|V|}\,|O_{p1}|^2 \, ,
	\label{eq:wp}
\end{align}
and $O_{p1}$ is the overlap integral~\eqref{eq:Z_overlap_integral}.
Equation~\eqref{eq:sumrule} follows from the Helmholtz
decomposition~\eqref{eq:HelmHoltz_decomp_V}: by
eq.~\eqref{eq:Tellegen_th}, $\Hvec^V_\Mvec[\varphivec_1]$ is orthogonal to all
the cavity fields $\hvec_p$, so that $\Bvec^V_\Mvec[\varphivec_1]$ is the
projection of $\mu_0\varphivec_1$ onto the space spanned by the $\hvec_p$.
The fields $\hvec_p$ are complete in the space of solenoidal fields with
vanishing normal component on $\partial V$ and orthonormal with respect to the
scalar product
$\langle\mathbf u,\mathbf v\rangle=\frac{1}{|V|}\int_V\mathbf u^*\cdot\mathbf v\,dV$.
Parseval's equality then gives
$\sum_pw_p=(\gamma\mu_0\Msat/|V_\text{M}|)\int_{V_\text{M}}
\varphivec_1^*\cdot\Bvec^V_\Mvec[\varphivec_1]/\mu_0\,dV$, and the
decomposition
$\Bvec^V_\Mvec[\varphivec_1]=\Bvec_\Mvec[\varphivec_1]+\mu_0\Hvec^\Delta_{1\Mvec}[\varphivec_1]$
yields $\omega_Q-\omega_\Delta$.

On the basis of the definitions above, the $\Avec$-formulation Hamiltonian,
written in terms of the mode amplitudes, reads
\begin{align}
	\nonumber
\mathcal{H}_\Avec={}&\hbar \omega_1|b|^2+\sum_{p\le P}\hbar \widetilde\omega_p|a_p|^2
	+\hbar \omega_Q|b|^2\\
	&-\sum_{p\le P}\hbar \left(G_p^*b+G_pb^*\right)(a_p+a_p^*) \, .
	\label{eq:HA}
\end{align}
The four terms are: the free magnetic mode; the free cavity modes (only the
first $P$ are retained); the normal part of the quadratic interaction term;
and the bilinear interaction between the magnetization and the cavity field.

The equations of motion are
\begin{equation}
	i\hbar\,\dot b=\frac{\partial \mathcal{H}_\Avec}{\partial b^*},\qquad
	i\hbar\,\dot a_p=\frac{\partial \mathcal{H}_\Avec}{\partial a_p^*} \, ,
	\label{eq:edom_modes_A}
\end{equation}
together with their complex conjugates. To find the eigenmodes of the coupled
system, we treat $b$, $b^*$, $a_p$, $a_p^*$ as independent variables, as in
Sec.~\ref{sec:finite_dim_NM_L_and_H}, and seek nontrivial solutions of
eq.~\eqref{eq:edom_modes_A} and of its complex conjugate in which all of them
vary in time as $e^{-i\Omega t}$. Eliminating the cavity amplitudes, neglecting
the coupling between $b$ and $b^*$ produced by the counter-rotating terms, which
shifts the eigenfrequencies by about $0.01\,\mathrm{MHz}$, and using
eq.~\eqref{eq:wp} together with the identity
$2\widetilde\omega_p/(\widetilde\omega_p^2-\Omega^2)
=2/\widetilde\omega_p+(2/\widetilde\omega_p)\,\Omega^2/(\widetilde\omega_p^2-\Omega^2)$,
we obtain the secular equation
\begin{align}
	\Omega=\omega_1+\Big(\omega_Q-\sum_{p\le P}w_p\Big)
	-\sum_{p\le P}w_p\,\frac{\Omega^2}{\widetilde\omega_p^2-\Omega^2} \, .
	\label{eq:A_secular}
\end{align}
The term in brackets is the static part: the static response of the retained
modes compensates the quadratic term $\omega_Q$, completely only for
$P\to\infty$ by the sum rule~\eqref{eq:sumrule}. For finite $P$ it equals
$\omega_\Delta+\sum_{p>P}w_p$, and only the fraction
$\eta(P)=\sum_{p\le P}w_p/\omega_Q$ is compensated (Table~\ref{tab:eta_P}).
With few modes the magnetic mode is shifted by almost the whole
$\omega_Q/2\pi\simeq5\,\mathrm{GHz}$, and the avoided crossing with the
TE$_{101}$ mode moves from $319.2\,\mathrm{mT}$ to $141.7\,\mathrm{mT}$ for
$P=1$, $153.7\,\mathrm{mT}$ for $P\simeq10^6$, and $255.1\,\mathrm{mT}$ for
$P\simeq10^9$. In the formulation of Sec.~\ref{sec:L_and_H_of_the_coupled_problem}
the static part is absent by construction, and including all the other cavity
modes changes the polariton frequencies obtained with the TE$_{101}$ mode alone
by about $0.4\,\mathrm{MHz}$.

\begin{table}[h]
	\caption{Compensated fraction $\eta(P)$, uncompensated static part
		$(\omega_Q-\sum_{p\le P}w_p)/2\pi$, and roots $\Omega/2\pi$ of
		eq.~\eqref{eq:A_secular} at $\mu_0H_\mathrm{dc}=319.1\,\mathrm{mT}$, as
		functions of the number $P$ of retained cavity modes
		($x_c=2\,\mathrm{mm}$). For finite $P$ only the root of magnetic character
		is reported (the other $P$ roots lie at the cavity frequencies); for
		$P\to\infty$ the two roots closest to $\widetilde\omega_1$ are reported,
		which are fully hybridized.}
	\label{tab:eta_P}
	\centering
	\begin{tabular}{rccc}
		\hline\hline
		$P$ & $\eta(P)$ & static part (GHz) & $\Omega/2\pi$ (GHz)\\
		\hline
		$1$      & $1\times10^{-5}$ & $4.97$  & $8.968$\\
		$10^6$   & $0.067$          & $4.64$  & $8.630$\\
		$10^8$   & $0.326$          & $3.35$  & $7.342$\\
		$10^9$   & $0.638$          & $1.80$  & $5.793$\\
		$\infty$ & $1$              & $0.002$ & $3.989$, $4.009$\\
		\hline\hline
	\end{tabular}
\end{table}

\section{Alternative electric vector potential formulation}
\label{sec:C_formulation_polartiton}
Here we discuss a formulation based on the electric vector potential $\Cvec_1$,
introduced in Sec.~\ref{sec:electric_C}, in which the interaction Lagrangian
between the cavity field and the magnetization is chosen to be
eq.~\eqref{eq:L_int_1st_0}, rather than the interaction term appearing in the
Lagrangian \eqref{eq:cavity_fild_lagrangian_coup_0_f}.

With this choice, the conjugate momenta become
\begin{align}
	\label{eq:conj_momentum_C1_b}
	& \bm{\Pi}_{\Cvec_1}
	= \frac{\delta L}{\delta (\partial_t \Cvec_1)}
	= \mu_0 \partial_t \Cvec_1
	-\Bvec_\Mvec^V \, , \\[3pt]
	\label{eq:cnj_mom_M1_b}
	& \bm{\Pi}_{\Mvec_1}
	= \frac{\delta L}{\delta (\partial_t \Mvec_1)}
	= \frac{1}{2\gamma\Msat^2}
	\Mvec_0\times\Mvec_1 \, .
\end{align}

Taking into account
eq.~\eqref{eq:B1_in_terms_C1},
it follows from
eq.~\eqref{eq:conj_momentum_C1_b}
that
$\bm{\Pi}_{\Cvec_1}=-\Bvec_1$,
namely the canonical momentum coincides, up to the sign, with the total magnetic flux density,
including the magnetostatic contribution.

After straightforward algebraic manipulations, and the application of the Helmholtz
decomposition of $\Mvec_1$ (see Appendix~\ref{sec:Helmholtz_Decomposition}), the corresponding Hamiltonian
takes the form
\begin{widetext}
	\begin{align}
		\mathcal H =
		\int_V
		\left[
		\frac{(\bm \Pi_{\Cvec_1})^2}{2\mu_0}
		+
		\frac{1}{2\varepsilon_0}
		(\nabla\times\Cvec_1)^2
		\right]
		dV
		+
		\int_{V_\text{M}}
		\frac{1}{2}
		\Mvec_1\cdot\Bvec_{\Mvec}^V[\Mvec_1]
		\,dV
		+
		\int_{V_\text{M}}
		\bm\Pi_{\Cvec_1}
		\cdot\Mvec_1
		\,dV + 	\widetilde{\mathcal H}_\text{M}[\Mvec_1] \, ,
		\label{eq:cavity_fild_Hamiltonian_coup_0_f_0}
	\end{align}
\end{widetext}
where $	\widetilde{\mathcal H}_\text{M}[\Mvec_1]$ is given by eq.~\eqref{eq:Hamiltonian_lin_coupled_0_f_a_bis}.

Unlike in the formulation based on Lagrangian \eqref{eq:cavity_fild_lagrangian_coup_0_f}, 
the interaction between the cavity field and
the magnetization appears explicitly in the Hamiltonian through the bilinear
term involving $\bm{\Pi}_{\Cvec_1}$ and $\Mvec_1$.

The canonical equation governing the magnetization dynamics is obtained,
after appropriate algebra, as
\begin{align}
	\nonumber
	\partial_t\Mvec_1
	&=
	\\
	&
	\gamma\Mvec_0\times
	\left[
	\delta_{\Mvec_1}\mathcal{H}_\text{M}
	+
	\left(
	\bm\Pi_{\Cvec_1}
	+
	\Bvec_{\Mvec}[\Mvec_1]
	\right)
	\right]
	\, ,
	\label{eq:LL_with_PiC1}
\end{align}
where $\Bvec_{\Mvec}[\Mvec_1]$ is defined as in eq.~\eqref{eq:general_VLL_Gqs_lin_B}.

Equation \eqref{eq:LL_with_PiC1}
is again affected by the same cancellation problem discussed in
Appendix~\ref{Sec:vector_pot_form_polaritons}. Indeed, the quantity
$\bm\Pi_{\Cvec_1}+\Bvec_{\Mvec}[\Mvec_1] = -(\Bvec_1- \Bvec_{\Mvec}[\Mvec_1] )$
is usually relatively small, but it is obtained as the sum of two nearly opposite
quantities that are not necessarily small. Consequently, this formulation
also behaves poorly under finite modal truncation.

\section{Nonsingularity of the Faddeev--Jackiw matrix}
\label{sec:F_nonsingular}

Here we show that the matrix $F$ in eq.~\eqref{eq:F_FJ_matrix} is
nonsingular for any coupling matrix $\Gamma\in\mathbb{C}^{M\times P}$.
Define
\begin{align}
	\Sigma_n=
	\begin{pmatrix}
		0 & -i\mathbbm{I}_n\\
		i\mathbbm{I}_n & 0
	\end{pmatrix},
	\label{eq:Sigma_n}
\end{align}
where each block is $n\times n$. This matrix satisfies
$\Sigma_n^{-1}=\Sigma_n$ and $\det\Sigma_n=(-1)^n$.
We also define
\begin{align}
	Y=
	\begin{pmatrix}
		\Gamma^*\\
		\Gamma
	\end{pmatrix},
	\qquad
	R=
	\begin{pmatrix}
		\mathbbm{I}_P & \mathbbm{I}_P
	\end{pmatrix},
\end{align}
where $Y$ and $R$ have dimensions $2M\times P$ and
$P\times 2P$, respectively.

Grouping the magnetic coordinates into a block of size $2M$ and the
electromagnetic coordinates into a block of size $2P$,
eq.~\eqref{eq:F_FJ_matrix} can be written as
\begin{align}
	F=\hbar
	\begin{pmatrix}
		\Sigma_M & YR\\
		-R^TY^T & \Sigma_P
	\end{pmatrix}.
	\label{eq:F_compact}
\end{align}
Since $\Sigma_P^{-1}=\Sigma_P$, the Schur-complement formula
\cite{HornJohnson2013} gives
\begin{align}
	\det F
	=
	\hbar^{2(M+P)}
	\det\Sigma_P\,\det S,
	\label{eq:detF_Schur}
\end{align}
where the $2M\times2M$ Schur complement is
\begin{align}
	S
	=
	\Sigma_M
	+
	Y\bigl(R\Sigma_P R^T\bigr)Y^T .
\end{align}
A direct calculation gives
\begin{align}
	R\Sigma_P R^T
	=
	\begin{pmatrix}
		\mathbbm{I}_P & \mathbbm{I}_P
	\end{pmatrix}
	\begin{pmatrix}
		0 & -i\mathbbm{I}_P\\
		i\mathbbm{I}_P & 0
	\end{pmatrix}
	\begin{pmatrix}
		\mathbbm{I}_P\\
		\mathbbm{I}_P
	\end{pmatrix}
	=
	0,
\end{align}
and therefore
\begin{align}
	S=\Sigma_M.
\end{align}
Using $\det\Sigma_n=(-1)^n$, we finally obtain
\begin{align}
	\det F
	=
	(-1)^{M+P}\,
	\hbar^{2(M+P)}
	\neq 0.
	\label{eq:detF}
\end{align}
Consequently, $F$ is nonsingular for any coupling matrix $\Gamma$,
without any assumption on the magnitude of the coupling coefficients.

\end{document}